\documentstyle[11pt,amssymb,epsf]{article}

\def\ben{\begin{equation}}
\def\een{\end{equation}}

\let\a=\alpha    
    
  \let\n=\nu

\let\C=\Chi

\def\nn{\nonumber} \def\bd{\begin{document}} \def\ed{\end{document}}
\def\ds{\documentstyle} \let\fr=\frac \let\bl=\bigl \let\br=\bigr
\let\Br=\Bigr \let\Bl=\Bigl
\let\bm=\bibitem
\let\na=\nabla
\let\pa=\partial \let\ov=\overline
\newcommand{\be}{\begin{equation}}
\newcommand{\ee}{\end{equation}}
\def\ba{\begin{array}}
\def\ea{\end{array}}
\def\ft#1#2{{\textstyle{{\scriptstyle #1}\over {\scriptstyle #2}}}}
\def\fft#1#2{{#1 \over #2}}
\def\del{\partial}
\def\vp{\varphi}
\def\sst#1{{\scriptscriptstyle #1}}
\def\oneone{\rlap 1\mkern4mu{\rm l}}
\def\td{\tilde}
\def\wtd{\widetilde}
\def\ie{\rm i.e.\ }
\def\dalemb#1#2{{\vbox{\hrule height .#2pt
        \hbox{\vrule width.#2pt height#1pt \kern#1pt
                \vrule width.#2pt}
        \hrule height.#2pt}}}
\def\square{\mathord{\dalemb{6.8}{7}\hbox{\hskip1pt}}}
\newcommand{\ho}[1]{$\, ^{#1}$}
\newcommand{\hoch}[1]{$\, ^{#1}$}
\newcommand{\bea}{\begin{eqnarray}}
\newcommand{\eea}{\end{eqnarray}}
\newcommand{\ra}{\rightarrow}
\newcommand{\lra}{\longrightarrow}
\newcommand{\Lra}{\Leftrightarrow}
\newcommand{\ap}{\alpha^\prime}
\newcommand{\bp}{\tilde \beta^\prime}
\newcommand{\tr}{{\rm tr} }
\newcommand{\Tr}{{\rm Tr} }
\def\0{{\sst{(0)}}}
\def\1{{\sst{(1)}}}
\def\2{{\sst{(2)}}}
\def\3{{\sst{(3)}}}
\def\4{{\sst{(4)}}}
\def\5{{\sst{(5)}}}
\def\6{{\sst{(6)}}}
\def\7{{\sst{(7)}}}
\def\8{{\sst{(8)}}}
\def\n{{\sst{(n)}}}
\def\cA{{{\cal A}}}
\def\cF{{{\cal F}}}
\def\tV{\widetilde V}
\def\tW{\widetilde W}
\def\tH{\widetilde H}
\def\tE{\widetilde E}
\def\tF{\widetilde F}
\def\tA{\widetilde A}
\def\im{{{\rm i}}}
\def\tY{{{\wtd Y}}}
\def\ep{{\epsilon}}
\def\vep{{\varepsilon}}
\def\R{\rlap{\rm I}\mkern3mu{\rm R}}
\def\bD{{{\bar D}}}
\def\cD{{{\cal D}}}

\def\R{\rlap{\rm I}\mkern3mu{\rm R}}
\def\bD{{{\bar D}}}
\def\R{{{\Bbb R}}}
\def\C{{{\Bbb C}}}
\def\H{{{\Bbb H}}}
\def\CP{{{\Bbb C}{\Bbb P}}}
\def\RP{{{\Bbb R}{\Bbb P}}}
\def\Z{{{\Bbb Z}}}
\def\bA{{{\Bbb A}}}
\def\bB{{{\Bbb B}}}
\def\bC{{{\Bbb C}}}
\def\bR{{{\Bbb R}}}
\def\bD{{{\Bbb D}}}
\def\bE{{{\Bbb E}}}
\def\bZ{{{\Bbb Z}}}
\def\Re{{{\frak{Re}}}}
\def\Im{{{\frak{Im}}}}
\def\cosec{{\,\hbox{cosec}\,}}
\def\Gm{{\Gamma_{\!\! -}}}
\def\Gp{{\Gamma_{\!\! +}}}
\def\stan{{standard }}
\def\nonstan{{supernumerary }}

\def\cosech{{\hbox{cosech}}}

\def\etcyc{{\hbox{and cyclic}}}
\def\btheta{{\bar\theta}}

\newcommand{\tamphys}{\it Center for Theoretical Physics,
Texas A\&M University, College Station, TX 77843, USA}
\newcommand{\umich}{\it Michigan Center for Theoretical Physics,
University of Michigan\\ Ann Arbor, MI 48109--1120, USA}
\newcommand{\princeton}{\it Department of Physics,
Princeton University\\ Princeton, NJ 08544, USA}
\newcommand{\upenn}{\it Department of Physics and Astronomy,\\
University of Pennsylvania, Philadelphia,  PA 19104, USA}
\newcommand{\SISSA}{\it  SISSA-ISAS and INFN, Sezione di Trieste\\
Via Beirut 2-4, I-34013, Trieste, Italy}

\newcommand{\mitchell}{\it George P. and Cynthia W. 
Mitchell Institute for Fundamental Physics,\\
Texas A\&M University, College Station, TX 77843-4242, USA}

\newcommand{\newton}{\it Isaac Newton Institute for Mathematical Sciences,\\
20 Clarkson Road,  University of Cambridge,
Cambridge CB3 0EH, UK}

\newcommand{\ihp}{\it Institut Henri Poincar\'e\\
  11 rue Pierre et Marie Curie, F 75231 Paris Cedex 05}

\newcommand{\damtp}{\it DAMTP, Centre for Mathematical Sciences,
 Cambridge University\\  Wilberforce Road, Cambridge CB3 OWA, UK}
\newcommand{\itp}{\it Institute for Theoretical Physics, University of
California\\ Santa Barbara, CA 93106, USA}

\newcommand{\auth}{
James T. Liu\hoch{\dagger1}, H. L\"u\hoch{\star2} 
and C.N. Pope\hoch{\star2}}

\begin{document}
\begin{flushright}
\hfill{MCTP-02-64}\\
\hfill{MIFP-02-07}\\
\hfill{PUPT-2066}\\
\hfill{\bf hep-th/0212037}
\end{flushright} 

\begin{center}  

{\large {\bf The Radion Mode in Consistent Brane-World Reductions}}   

\vspace{15pt}

\auth

\vspace{7pt}
{\hoch{\dagger}\umich}

\vspace{7pt}
{\hoch{\star}\mitchell}

\vspace{7pt}
{\hoch{\dagger}\princeton}

\underline{ABSTRACT}
\end{center}  

   We construct consistent brane-world Kaluza-Klein reductions involving
the radion mode that measures the separation of the domain-wall branes.  
In these new examples, we can obtain matter supermultiplets coupled
to supergravity on the brane, starting from pure gauged supergravity in 
the higher dimension.  This contrasts with previously-known examples
of consistent brane-world reductions involving the radion, where 
either pure supergravity reduced to pure supergravity, or else 
supergravity plus matter reduced to supergravity plus matter.  As well
as considering supersymmetric reductions, we also show that there exist
broader classes of consistent reductions of bosonic systems. These include
examples where the lower-dimensional theory has non-abelian Yang-Mills
fields and yet the scalar sector has a potential that admits Minkowski
spacetime as a solution.  Combined with a sphere reduction to obtain the
starting point for the brane-world reduction, this provides a Kaluza-Klein
mechanism for obtaining non-abelian gauge symmetries from the geometry of 
the reduction, whilst still permitting a Minkowski vacuum in the
lower dimension.

{\vfill\leftline{}\vfill
\vskip 10pt \footnoterule
{\footnotesize \hoch{1}
Research supported in part by DOE grant DE-FG02-95ER40899
\vskip  -12pt} \vskip   14pt
{\footnotesize \hoch{2}
Research supported in part by DOE grant DE-FG03-95ER40917
\vskip -12pt} \vskip 14pt
}

\pagebreak
\setcounter{page}{1}

\tableofcontents
\addtocontents{toc}{\protect\setcounter{tocdepth}{2}}
\newpage 

\section{Introduction}

   In conventional Kaluza-Klein reductions, the internal space is
taken to be compact with isometries, leading to
lower-dimensional theories that comprise a finite number of massless
modes, together with infinite towers of massive modes.  In certain
circumstances, it is possible to perform a consistent truncation of
all the massive modes (meaning that setting the massive modes to zero
is consistent with their equations of motion).  In some cases, such as
Kaluza-Klein reduction on a circle, torus, or other group manifold,
there is a clear-cut group-theoretic reason for the consistency of the
truncation, in that all the fields that are group singlets are
retained, whilst all those that are non-singlets are discarded.
Clearly, then, the retained fields cannot act as sources for those
that are set to zero.  In other cases, there are much more remarkable
consistent reductions for which there is no fully-understood
group-theoretic explanation.  Examples of this kind include the
reductions of $D=11$ supergravity on $S^4$ or $S^7$, and the reduction
of type IIB supergravity on $S^5$.

   In the usual circle reduction from $(D+1)$ to $D$ dimensions,
retaining just the massless sector, the internal space (and the
reduction ansatz) is assumed to have a $U(1)$ isometry.  A rather
different kind of reduction has been considered recently, in which the
$D$-dimensional world is viewed as the world-volume of a $(D-1)$-brane
(\ie a domain wall) embedded in $(D+1)$ dimensions.  Clearly, the
one-dimensional transverse space no longer has a $U(1)$ isometry; it
is broken by the location of the domain wall, and by the warp-factor
of the domain-wall metric.  Thus the conventional Kaluza-Klein
technique has to be modified for these new situations.

     Firstly, it is necessary to examine whether gravity would indeed
localise on the brane, so that one has a genuinely lower-dimensional
theory.  One way to achieve the localisation is by considering two
branes, one located at each end of a finite interval.  In such a
scenario, the localisation of gravity is guaranteed, since the
internal space is finite, and so the massive Kaluza-Klein tower will
have a discrete mass spectrum.  An alternative to this
compactification is to make use of the fact that with the domain-wall
warp factor, it is possible to to trap gravity even when there is only
a single brane, with an extra dimension of infinite extent
\cite{rsII}.  This is because although the extra dimension is
infinite, its volume is finite owing to the warp factor.  Using either
the single-brane or double-brane scenario, it is possible to arrive at
a lower-dimensional gravity theory on the brane.  In fact, the
equations of motion for the domain wall solution require only that the
world-volume metric have vanishing Ricci-tensor, rather than the more
stringent condition of vanishing Riemann tensor and hence Minkowskian
spacetime.  Furthermore, the domain-wall solution preserves a certain
fraction of supersymmetry (typically $\ft12$).  It follows that one
would expect to find a supergravity theory with lesser supersymmetry
trapped on the domain-wall world-volume.  It has been shown in
\cite{bob1,bob2,dufliusab} that this can indeed be the case.  The
associated consistent reduction procedure is known as a brane-world
Kaluza-Klein reduction.

            In these examples, a pure supergravity theory was obtained
on the brane, by using a modified, but nonetheless consistent,
Kaluza-Klein procedure.  It is of considerable interest to see whether
matter supermultiplets can also arise through such brane-world
Kaluza-Klein reductions.  Generating matter using Kaluza-Klein is not
always guaranteed.  For example, in the Horava-Witten model, the
$E_8\times E_8$ Yang-Mills fields of the heterotic string are not
expected to come from a Kaluza-Klein reduction; rather, their
existence is argued on the grounds of anomaly cancellation \cite{hw}.
In the present paper, we shall consider examples where we can obtain
matter supermultiplets from consistent brane-world Kaluza-Klein
reductions.

    In the reductions that we shall obtain in this paper, the
breathing mode (\ie the scalar that measures the ``size'' of the extra
dimension) plays an important role.  It can also be thought of as a
``radion mode,'' since in the double-brane picture it measures the
relative separation of the two branes in the transverse dimension.  In
fact consistent brane-world Kaluza-Klein reductions involving the
radion mode were first introduced in \cite{bob2}.  In those examples,
which include the reduction of the massive IIA theory to $D=9$, and
the reduction of gauged $D=8$ pure supergravity to $D=7$, the radion
mode becomes the dilaton of the pure supergravity multiplet in the
lower dimension.  A further example of a consistent brane-world
reduction involving the radion mode was then obtained in
\cite{lehste}; in that case the starting point was gauged
five-dimensional ${\cal N}=2$ supergravity coupled to a hypermultiplet, and
the radion became the scalar member of a chiral matter multiplet in
four dimensional ${\cal N}=1$ supergravity.

    In this paper, we shall obtain various examples of consistent
brane-world reductions involving the radion, which give rise to
lower-dimensional supergravities (with a halving of supersymmetry)
coupled to matter multiplets.  We begin in section 2 with a general
discussion of the circumstances under which we can obtain a consistent
brane-world reduction of a bosonic theory comprising gravity, a
dilaton with an exponential potential, and a $p$-form field strength.
We find that a consistent reduction is possible if there is a specific
relation between the dilaton coupling to the $p$-form and the dilaton
coupling in the exponential potential.  We make extensive use of these
general results in the subsequent sections, when we consider
consistent reductions of gauged supergravities.  Our first
supergravity example, in section 3, starts from gauged ${\cal N}=2$
supergravity in $D=7$.  We show that a consistent brane-world
reduction is possible in which we obtain ungauged ${\cal N}=(1,0)$ chiral
supergravity in $D=6$, coupled to a chiral tensor multiplet.  The
radion mode in this reduction forms the scalar member of the chiral
tensor multiplet.  (The brane-world reduction to give pure ${\cal N}=(1,0)$
chiral six-dimensional supergravity was obtained in \cite{bob1}.)  Solutions
in the six-dimensional theory can then be lifted back to $D=7$.  We consider
the BPS dyonic string, and demonstrate that in $D=7$ it leads to a 
bending of the domain walls.

   In section 4, we obtain further supersymmetric consistent
brane-world reductions to ungauged supergravities plus matter.  In one
of these, we obtain five-dimensional supergravity with a vector
multiplet, starting from six-dimensional gauged ${\cal N}=(1,1)$
supergravity.  In fact this, and the above reduction from $D=7$, are
the first examples where supermatter as well as supergravity is
obtained in consistent brane-world reductions of pure supergravity
theories.  In a further example, we obtain four-dimensional ${\cal N}=1$
supergravity with a chiral multiplet, starting from gauged $D=5$
supergravity with a single vector multiplet.

   In section 5 we consider some extended bosonic systems for which we
can obtain consistent brane-world reductions.  Of particular interest
are cases where the starting point is the bosonic sector of a gauged
supergravity in which we now augment the previous brane-world
reductions by including $SU(2)$ Yang-Mills fields.  We find that
consistent brane-world reductions are possible in which we end up with
these Yang-Mills fields in the lower dimension, but still in a theory
where there is no cosmological term or scalar potential.  The
higher-dimensional gauged theories can themselves be obtained by $S^3$
reduction of ungauged supergravity in a yet higher dimension. We
therefore have the intriguing situation that we can view the $S^3$
plus brane-world reduction as a $(3+1)$-dimensional reduction scheme
in which non-abelian Yang-Mills emerges from Kaluza-Klein reduction,
without any cosmological term or scalar potential being generated.
The $(3+1)$-dimensional reduction can viewed as a reduction on a
singular cone over $S^3$.

   In section 6 we consider the brane-world reductions of bosonic
theories with additional scalars as well as the dilaton in the higher
dimension.  Such theories typically arise as subsectors of gauged
supergravities.  We find that under appropriate circumstances we can
obtain consistent brane-world reductions in which all the extra
scalars are retained.  The scalar potential in the lower dimension is
related to that in the higher dimension, but with a modification which
means, in particular, that it admits a Minkowski spacetime vacuum.

\section{Consistent Reduction of a $p$-form and Radion}

    We begin by deriving a general result for a consistent
Kaluza-Klein brane-world reduction of a $(D+1)$-dimensional theory
comprising a metric, a dilatonic scalar and a $p$-form field strength.
In the class of theory we shall be considering, the dilaton has a
scalar potential which is a single exponential function:
\be
\hat {\cal L} = \hat R\, {\hat*\oneone} - \ft12 {\hat *d\hat\phi}\wedge
d\hat\phi - \ft12 e^{\gamma\, \hat\phi}\, {\hat * \hat F}\wedge
\hat F + g^2\, e^{a\,\hat\phi}\, {\hat *\oneone}\,.\label{d1lag}
\ee
It will frequently turn out to be convenient to parameterise the constant 
$a$ in the scalar potential in terms of a quantity $\Delta$, 
where\footnote{$\Delta$ is preserved under toroidal dimensional reduction,
as discussed in \cite{stainless}.}
\be
a^2 \equiv  \Delta + \fft{2D}{D-1}= \Delta + 2 + \fft{2}{D-1}
\,.\label{avalue}
\ee

  The equations of motion for this theory,
\bea
\hat R_{AB} &=& \ft12 \del_A\hat\phi\,\del_B\hat\phi + 
\fft1{2\, (p-1)!}\, \Big( \hat F^2_{AB} - \fft{(p-1)}{p\, (D-2)}\, 
\hat F^2\, \eta_{AB}\Big) -\fft{g^2}{D-1}\, e^{a\hat\phi}\, \eta_{AB}\,,
\nn\\
\hat{\square} \hat\phi &=& \fft{\gamma}{2 p!}\, e^{\gamma\, \hat\phi}\,
\hat F^2 - a\, g^2\, e^{a\hat\phi}\,,\label{d1eom}\\
d(e^{a\hat\phi} \, {\hat * \hat F})&=&0\,,\nn
\eea
do not admit an AdS$_{D+1}$ ``vacuum'' solution, but they do allow a 
domain wall, given by
\bea
d\hat s^2 &=& W^{\ft{4}{(D-1)\, \Delta}}\, dx^\mu\, dx_\mu  + 
W^{\ft{4D}{(D-1)\, \Delta}}\, dy^2\,,\nn\\
e^{\hat \phi} &=& 
W^{-\ft{2a}{\Delta}}\,,\qquad \hat F=0\,,\label{domainwall}
\eea
where $W$ is a linear function of the transverse coordinate $y$;
\be
W = 1 + m\, y\,,\qquad m^2 = -\ft12 \Delta\, g^2\,.
\ee
(For an actual domain wall one would replace $y$ by $|y|$.  Since our
focus here is on the Kaluza-Klein reductions rather than the
properties of the wall itself, it is preferable for our present
purposes to omit the modulus sign, which would lead to additional
delta-function contributions in the curvature.)

    The domain-wall ``vacuum'' can be thought of as a background
solution around which a Kaluza-Klein brane-world reduction can be
performed.  In fact, we can do much better than merely describing
linearised fluctuations; in appropriate circumstances we can obtain a
fully consistent Kaluza-Klein embedding that is exact to all
non-linear orders.  Specifically, we find that we can obtain a 
consistent reduction if the constants
$\gamma$ and $a$ in (\ref{d1lag}) are related by
\be
\gamma= -\fft{2(p-1)}{(D-1)\,a}\,.\label{gamma-a}
\ee
We find that the reduction ansatz is given by\footnote{Here, and
throughout the paper, we shall place hats on the higher-dimensional 
fields, which depend, {\it a priori}, on the lower-dimensional coordinates
$x^\mu$ and the extra coordinate $y$.  Unhatted fields live in the 
lower dimension, and depend only on the $x^\mu$ coordinates.} 
\bea
d\hat s^2 &=& W^{\ft{4}{(D-1)\, \Delta}}\, e^{2\alpha\, \varphi}\, 
ds^2 + 
W^{\ft{4D}{(D-1)\, \Delta}}\, e^{-2(D-2)\, \alpha\, \varphi}\, 
dy^2\,,\nn\\
e^{a\,\hat \phi} &=& 
W^{-\ft{2 a^2}{\Delta}}\, e^{2(D-2)\, \alpha\, \varphi}
\,,\qquad \hat F=F\,.\label{d1drred}
\eea
It is useful to record that in the obvious vielbein basis $\hat e^a
=W^{2/((D-1)\, \Delta)}\, e^{\alpha\, \varphi}\, e^a$, $\hat e^0 = 
W^{2D/((D-1)\, \Delta)}\, e^{-(D-2)\, \alpha\, \varphi}\, dy$, the 
torsion-free spin connection and the 
components of the Ricci tensor turn out to be given by
\bea
\hat \omega_{0a} &=& -(D-2)\, \alpha\, W^{-\ft2{(D-1)\, \Delta}}\, 
e^{-\alpha\, \varphi}\, \hat e^0 - \fft{2m}{(D-1)\, \Delta}\, 
W^{\ft{2D + (D-1)\, \Delta}{(D-1)\, \Delta}}\, 
e^{(D-2)\, \alpha\, \varphi}\, \hat e^a\,,\nn\\
\hat \omega_{ab} &=& \omega_{ab} + \alpha\, W^{-\ft2{(D-1)\, \Delta}}\,
e^{-\alpha\, \varphi}\, (\del_b\varphi\, \hat e^a - \del_a\varphi\, 
\hat e^b)\,,\nn\\
\hat R_{00} &=& (D-2)\, \alpha\, e^{-2\alpha\, \varphi}\, W^{\ft2{D-1}}\, 
\square\varphi\,,\nn\\
\hat R_{0a} &=& m\, \alpha\, (D-2)\, W^{\ft2{D-1}}\, 
e^{(D-3)\, \alpha\, \varphi}\, \del_a\, \varphi\,,\\
\hat R_{ab} &=& W^{\ft2{D-1}}\, e^{-2\alpha\, \varphi}\, \Big(
R_{ab} - (D-1)(D-2)\, \alpha^2\, \del_a\varphi\, \del_b\varphi 
-\alpha\, \square\varphi\, \eta_{ab}\Big) \nn\\
&& -\fft{m^2}{D-1} \, W^{\ft2{D-1}}\, e^{2(D-2)\, \alpha\, \varphi}\, 
\eta_{ab}\,.\nn
\eea

   Substituting (\ref{d1drred}) into the higher-dimensional equations
of motion (\ref{d1eom}), we obtain $D$-dimensional equations of motion
for the metric, the $p$-form $F$ and the radion $\varphi$, which can
be derived from the Lagrangian
\be
{\cal L} = R\, {*\oneone} -\ft12 {*d\varphi}\wedge d\varphi - 
  \ft12 e^{-\ft{2(p-1)\,(\Delta +4)\,\alpha}{a^2}\,
\varphi}\, {*F}\wedge F\,,\label{dlag-1}
\ee
where $F=dA$ and we have chosen 
\be
\a^2 = \fft{a^2}{2(D-2)\, (D-1)\, (\Delta + 4)}\,,\label{alphaval}
\ee
so that the radion is canonically normalised.  Note that the relation
(\ref{gamma-a}) between $\gamma$ and $a$ is essential in order that
the $y$-dependence in the various higher-dimensional equations of motion
balances properly, giving rise to consistent lower-dimensional equations
of motion.

    In the above consistent reduction, a $p$-form in $(D+1)$
dimensions is reduced only to a $p$-form in $D$ dimensions.  Since a
$p$-form is dual to a $(D+1-p)$ form in the original
$(D+1)$-dimensional theory, there is dual description in terms of a
$\td p=D+1-p$ form field $\hat G\equiv e^{\gamma\, \hat\phi}\, {\hat *
\hat F}$, for which the reduction ansatz is
\be
 \hat G =W^{\ft{4}{\Delta}}\, G\wedge dy\,,
\ee
where the lower-dimensional field strength $G$ is a $(\td p-1)$-form.  This
field is related by dualisation to the previous $p$-form field $F$ in 
$D$ dimensions in the usual way, namely
\be
G = e^{-\ft{2(p-1)\,(\Delta + 4)\,\alpha}{a^2}\, \varphi}\, {* F}\,.
\ee
Thus an equivalent statement about the circumstances under which a 
consistent braneworld reduction of (\ref{d1lag}) can be performed is
that $\gamma$ must be related to $a$ by
\be
\gamma=-\fft{2(p-1)}{(D-1)\, a}\,,\qquad \hbox{or}\qquad 
 \gamma =  \fft{2(D-p)}{(D-1)\, a}\,.\label{gamma-arels}
\ee
In the first case, the $p$-form field strength is reduced according to
$\hat F=F$, while in the second case the reduction of the $p$-form $\hat F$ 
is instead performed using  
\be
 \hat F =W^{\ft{4}{\Delta}}\, F\wedge dy\,.
\label{second-red}
\ee
In this case the resulting $D$-dimensional Lagrangian is given by
\be
{\cal L} = R\, {*\oneone} -\ft12 {*d\varphi}\wedge d\varphi - 
  \ft12 e^{\ft{2(D-p)\,(\Delta +4)\,\alpha}{a^2}\,
\varphi}\, {*F}\wedge F\,,\label{dlag-2}
\ee
where $\alpha$ is again given by (\ref{alphaval}), and now $F=dA$ is a 
$(p-1)$-form.

    The consistent brane-world reduction that we have derived here
makes essential use of the radion field $\varphi$ that characterises
the scale in the $y$ direction transverse to the lower-dimensional
spacetime.  Such brane-world reductions involving the radion mode were
first introduced in some of the consistent brane-world reductions
obtained \cite{bob2}.  One of these was a reduction of massive
type IIA supergravity to give ${\cal N}=1$ ungauged supergravity in $D=9$,
and the other was a reduction of gauged supergravity in $D=8$ to give
ungauged ${\cal N}=2$ supergravity in $D=7$.  A further example of a
consistent brane-world reduction involving the radion mode was
obtained in \cite{lehste}, where the five-dimensional theory resulting
from a generalised Calabi-Yau reduction from $D=11$ was further
reduced to give ${\cal N}=1$ supergravity plus a chiral scalar multiplet in
$D=4$.

   In subsequent sections, we shall make use of the results that we
have obtained here in order to construct consistent brane-world
reductions of various gauged supergravity theories.  It turns out that
for all the examples we shall consider, the scalar potential in the
higher-dimensional gauged theory is of the single exponential form
in (\ref{d1lag}), with the constant $a$ given by (\ref{avalue})
with $\Delta=-2$.  In these cases, it follows from (\ref{avalue})
and (\ref{gamma-a}) that we have a consistent reduction if either
\be
a^2 =\fft{2}{D-1}\,,\qquad \gamma= -(p-1)\, a\,,\qquad
\hbox{and} \qquad \hat A=A\,,
\ee
yielding the $D$-dimensional Lagrangian
\be
{\cal L} = R\, {*\oneone} -\ft12{*d\varphi}\wedge d\varphi 
-\ft12 e^{-2(p-1)(D-1)\, \alpha\, \varphi}\,
{*F\wedge F}\,,
\ee
where $F=dA$ is a $p$-form, or else if
\be
a^2 =\fft{2}{D-1}\,,\qquad \gamma= (D-p)\, a\,,\qquad
\hbox{and} \qquad \hat A=W^{-2}\, A\wedge dy\,,
\ee
yielding the $D$-dimensional Lagrangian
\be
{\cal L} = R\, {*\oneone} -\ft12{*d\varphi}\wedge d\varphi 
-\ft12 e^{2(D-p)(D-1)\, \alpha\, \varphi}\,
{*F\wedge F}\,,
\ee
where $F=dA$ is a $(p-1)$-form.  In each case the metric and dilaton
reduction ansatz is
\bea
d\hat s^2 &=& W^{-\ft{2}{D-1}}\, e^{2\alpha\, \varphi}\, ds^2 +
W^{-\ft{2D}{D-1}}\, e^{-2(D-2)\,\alpha\,\varphi}\, dy^2\,,\nn\\
e^{a\, \hat \phi} &=& W^{\ft2{D-1}}\, e^{2(D-2)\, \alpha\, \varphi}\,,
\eea
and 
\be
\alpha^2 = \fft1{2(D-2)\, (D-1)^2}\,.
\ee

\section{$D=7$ Reduced to $D=6$, ${\cal N}=(1,0)$ supergravity with matter}
\label{7to6red}

   In this section, we shall carry out in detail the consistent
brane-world reduction of a gauged ${\cal N}=2$ seven-dimensional
supergravity.\footnote{We use the convention where the allowed
supersymmetries in $D=7$ are ${\cal N}=2$ and ${\cal N}=4$.  Thus, the
${\cal N}=2$ theory has half of maximal supersymmetry.}  Gauged ${\cal
N}=4$ supergravity can be obtained via a consistent $S^4$ reduction
from $D=11$ \cite{vann1,vann2}, and the ${\cal N}=2$ theory can be
obtained as a truncation of this.  In the process, the $SO(5)$
Yang-Mills fields of the ${\cal N}=4$ theory are truncated to $SU(2)$.
Explicit expressions for the $S^4$ reduction were obtained in
\cite{lupo-d7}. The bosonic field content comprises the metric, a
dilaton $\hat\phi$, a 4-form field strength $\hat F_\4$, and the
$SU(2)$ Yang-Mills fields $\hat F_\2^i$.  There is a scalar potential
which is the sum of three different exponentials of the dilaton $\hat
\phi$, of the form
\be
V = 2 g_1^2\, e^{-\fft2{\sqrt{10}}\, \hat\phi} + 2g_1\, g_2\, 
e^{\fft{3}{\sqrt{10}}\,\hat\phi} - \ft14 g_2^2\, 
e^{\fft{8}{\sqrt{10}}\,\hat\phi}\,.\label{3termpot}
\ee
This has a stationary point, and hence the theory admits an AdS$_7$
``vacuum'' solution.  Note that the two constants $g_1$ and $g_2$ have
interpretations as the $SU(2)$ gauge coupling and a topological mass
term respectively.

   A consistent brane-world reduction that yielded just ungauged
chiral ${\cal N}=(1,0)$ supergravity in six dimensions was constructed in
\cite{bob1}.  It could be viewed as a fully non-linear generalisation
of a linearised Kaluza-Klein reduction around the AdS$_7$ vacuum.  In
the bosonic sector, the resulting six-dimensional theory comprised
just the metric and a self-dual 3-form.  In the present paper, we wish
to extend the scope of the brane-world reduction, so that we obtain
${\cal N}=(1,0)$ supergravity coupled to an ${\cal N}=(1,0)$ matter multiplet.
Specifically, we shall show how we can obtain the tensor matter
multiplet comprising an anti-self-dual 3-form plus a scalar field.  In
order to do this, we shall employ the reduction scheme derived in
section 2.  This reduction requires that there be only a single
exponential in the seven-dimensional theory, and that it be related to
the dilaton coupling for the 4-form $\hat F_\4$ in the specific way
discussed in section 2.  In fact we find that this can be achieved by
setting the topological mass term $g_2$ to zero.\footnote{The theory
with $g_2=0$ also arises as the Scherk-Schwarz group-manifold
reduction of ten-dimensional type I supergravity on $S^3$, truncated
to the pure ${\cal N}=2$ supergravity sector \cite{chamsab}. By contrast, the
theory where $g_1$ is instead set to zero arises from the generalised
reduction of $D=11$ supergravity on $T^4$ \cite{lupo-t4}.}

\subsection{The bosonic sector}
\label{7to6red1}

    After setting $g_2=0$ and relabeling $g_1=g$, the
seven-dimensional bosonic Lagrangian becomes
\bea 
\hat {\cal L}_7 &=& \hat R\, {\hat *\oneone} - \ft12 {\hat *
d\hat \phi}\wedge d\hat\phi - \ft12 e^{-\fft{4}{\sqrt{10}}\, \hat\phi}\, 
{\hat *\hat F_\4}\wedge \hat F_\4 -  \ft12 e^{\fft{2}{\sqrt{10}}\,
 \hat\phi}\,{\hat * \hat F_\2^i}\wedge \hat F_\2^i \nn\\
&&+ \ft12 \hat F_\2^i\wedge \hat F_\2^i \wedge \hat A_\3 + 2 g^2\, 
e^{-\fft{2}{\sqrt{10}}\, \hat\phi}\, {\hat *\oneone}\,,\label{d7lag-2}
\eea
where $\hat F_\4=d\hat A_\3$ and $\hat F_\2^i = d\hat A_\1^i + \ft12
g\, \ep_{ijk}\, \hat A_\1^j\wedge \hat A_\1^k$.  It is evident that
the 4-form dilaton coupling with $\gamma=-\fft4{\sqrt{10}}$ and the
scalar potential with $a=-\fft2{\sqrt{10}}$ satisfy the second of the
two criteria in (\ref{gamma-arels}), implying that we can obtain a
consistent reduction of the 4-form $\hat F_\4$ to give a 3-form in six
dimensions.\footnote{Note that the constant $a$ is given by
$\Delta=-2$ in (\ref{avalue}).  In fact in all our examples, the
strength of dilaton coupling in the scalar potential is characterised
by $\Delta=-2$.}  If we dualise the 4-form field strength to a 3-form,
the resulting Lagrangian can be obtained from the $SU(2)$
Scherk-Schwarz reduction of ${\cal N}=1$ supergravity in $D=10$
\cite{chamsab}.

    From the formulae in section 2, we are therefore led to the
following brane-world reduction ansatz for the seven-dimensional
theory:
\bea
d\hat s_7^2 &=& W^{-\fft25}\, e^{2\alpha\, \varphi}\, ds_6^2 
+ W^{-\fft{12}{5}}\, e^{-8\alpha\, \varphi}\, dy^2\,,\nn\\
e^{-\fft2{\sqrt{10}}\,\hat\phi} &=& W^{\fft25}\, e^{8\alpha\, \varphi}\,,
\label{7to6ans}
\\
\hat A_\3 &=& W^{-2}\, A_\2\wedge dy\,,\qquad \hat A_\1^i=0\,.\nn
\eea
where $\alpha=-1/(10\sqrt2)$, $W=1+m\, y$, and $m^2 = 2g^2$.
Substituting this into the equations of motion following from
(\ref{d7lag-2}), we find that we obtain a consistent Kaluza-Klein
reduction, resulting in six-dimensional equations that can be derived
from the Lagrangian
\be
{\cal L}_6 = R\, {*\oneone} - \ft12 {*d\varphi}\wedge d\varphi 
- \ft12 e^{-\sqrt2\, \varphi}\, {*F_\3}\wedge F_\3\,,\label{d6lag-1}
\ee
where $F_\3=dA_\2$.  This is precisely the bosonic sector of
six-dimensional ${\cal N}=(1,0)$ supergravity coupled to an ${\cal
N}=(1,0)$ tensor matter multiplet.  The supergravity multiplet
comprises the metric and the self-dual part of $F_\3$, and the tensor
multiplet comprises the ``radion'' $\varphi$ and the anti-self-dual
part of $F_\3$.

     In the next subsection, we shall show that the consistent
reduction we have performed here can be extended to include the
fermionic sector, and thus that we can obtain the full ${\cal N}=(1,0)$
supergravity coupled to the tensor multiplet, via the brane-world
reduction.

\subsection{The fermionic sector}

The bosonic Lagrangian (\ref{d7lag-2}) of the previous subsection has a
supersymmetric completion \cite{Townsend:1983kk}, given up to quartic
terms in the fermions by
\begin{eqnarray}
\hat{\cal L}_7^{\rm fermion}&=&-\ft12\bar{\hat\psi}{}_M^i
\hat\gamma^{MNP}\hat D_N^{\vphantom{i}}\hat\psi_{P\,i}^{\vphantom{i}}
-\ft12\bar{\hat\lambda}{}^i\hat\gamma^M\hat
D_M^{\vphantom{i}}\hat\lambda_i^{\vphantom{i}}\nonumber\\
&&-\ft1{16}[\ft1{12}\bar{\hat\psi}{}_M^i\hat\gamma^{MNABCD}
\hat\psi_{N\,i}^{\vphantom{i}}+\bar{\hat\psi}{}^{A\,i}
\hat\gamma^{BC}\hat\psi_i^D]e^{-\fft2{\sqrt{10}}\hat\phi}
\hat F_{ABCD}\nonumber\\
&&-\ft{i}{4\sqrt{2}}[\ft12\bar{\hat\psi}{}_M^i
\hat\gamma^{MNAB}\hat\psi_{N\,j}^{\vphantom{i}}+\bar{\hat\psi}{}^{A\,i}
\hat\psi_j^B]e^{\fft1{\sqrt{10}}\hat\phi}\hat F_{AB\,i}{}^j\nonumber\\
&&+\ft1{48\sqrt{5}}(\bar{\hat\lambda}{}^i\hat\gamma^M\hat\gamma^{ABCD}
\hat\psi_{M\,i}^{\vphantom{i}})e^{-\fft2{\sqrt{10}}\hat\phi}\hat F_{ABCD}
\nonumber\\
&&-\ft{i}{4\sqrt{10}}(\hat{\bar\lambda}{}^i\hat\gamma^M\hat\gamma^{AB}
\hat\psi_{M\,j}^{\vphantom{i}})e^{\fft1{\sqrt{10}}\hat\phi}\hat
F_{AB\,i}{}^j -\ft1{2\sqrt{2}}(\bar{\hat\lambda}{}^i\hat\gamma^M
\hat\gamma^A\hat\psi_{M\,i}^{\vphantom{i}})\partial_A\hat\phi\nonumber\\
&&+\ft1{320}(\bar{\hat\lambda}{}^i\hat\gamma^{ABCD}
\hat\lambda_i^{\vphantom{i}})e^{-\fft2{\sqrt{10}}\hat\phi}\hat
F_{ABCD}-\ft{3i}{40\sqrt{2}}(\bar{\hat\lambda}{}^i
\hat\gamma^{AB}\hat\lambda_j^{\vphantom{i}})e^{\fft1{\sqrt{10}}\hat\phi}\hat
F_{AB\,i}{}^j\nonumber\\
&&-\ft14ge^{-\fft1{\sqrt{10}}\hat\phi}\bar{\hat\psi}{}_\mu^i
\hat\gamma^{\mu\nu}\hat\psi_{\nu\,i}^{\vphantom{i}}-\ft1{2\sqrt{5}}g
e^{-\fft1{\sqrt{10}}\hat\phi}\bar{\hat\psi}{}_\mu^i\hat\gamma^\mu
\hat\lambda_i^{\vphantom{i}}-\ft3{20}g
e^{-\fft1{\sqrt{10}}\hat\phi}\bar{\hat\lambda}{}^i
\hat\lambda_i^{\vphantom{i}}.
\label{eq:d7flag}
\end{eqnarray}
The fully gauge-covariant derivative $\hat D_\mu$ is defined
as, {\it e.g.}
\begin{equation}
\hat D_M\hat\epsilon_i=\hat\nabla_M\hat\epsilon_i+\ft{i}2g\hat A_{M\,i}{}^j
\hat\epsilon_j
\end{equation}
where $\hat A_{M\,i}{}^j$ is given by
$\hat A_{M\,i}{}^j\equiv \hat A_M^k(-\sigma^k)_i{}^j$.  This results in
a field strength given by
$\hat F_{MN\,i}{}^j=\partial_M\hat A_{N\,i}{}^j+\fft{i}2g
\hat A_{M\,i}{}^k\hat A_{N\,k}{}^j-({\scriptstyle M}\leftrightarrow
{\scriptstyle N})$.  The supersymmetry transformations are given by
\cite{Townsend:1983kk}
\begin{eqnarray}
\delta\hat\psi_{M\,i}&=&[\hat D_M+\ft1{160}(\hat\gamma_M{}^{NPQR}
-\ft83\delta_M^N\hat\gamma^{PQR})e^{-\fft2{\sqrt{10}}\hat\phi}
\hat F_{NPQR}-\ft1{5\sqrt{2}}g\hat\gamma_Me^{-\fft1{\sqrt{10}}\hat\phi}]
\hat\epsilon_i\nonumber\\
&&\qquad+[\ft{i}{20\sqrt{2}}(\hat\gamma_M{}^{NP}-8\delta_M^N\hat\gamma^P)
e^{\fft1{\sqrt{10}}\hat\phi}\hat F_{NP\,i}{}^j]\hat\epsilon_j,\nonumber\\
\delta\hat\lambda_i&=&[-\ft1{2\sqrt{2}}\hat\gamma^M\partial_M\hat\phi
+\ft1{48\sqrt{5}}e^{-\fft2{\sqrt{10}}\hat\phi}\hat
F_{MNPQ}\hat\gamma^{MNPQ}+\ft1{\sqrt{10}}ge^{-\fft1{\sqrt{10}}\hat\phi}]
\hat\epsilon_i\nonumber\\
&&\qquad+[-\ft{i}{4\sqrt{10}}e^{\fft1{\sqrt{10}}\hat\phi}\hat
F_{MN\,i}{}^j\hat\gamma^{MN}]\hat\epsilon_j,
\label{eq:7fsusy}
\end{eqnarray}
for the fermions, and
\begin{eqnarray}
\delta\hat\phi&=&-\ft1{2\sqrt{2}}\bar{\hat\epsilon}{}^i\hat\lambda_i\,,\nn\\
\delta\hat e_M^A&=&\ft14\bar{\hat\epsilon}{}^i\gamma^A\hat\psi_{M\,i}\,,
\nonumber\\
\delta\hat A_{MNP}&=&e^{\fft2{\sqrt{10}}\hat\phi}[\ft34\bar{\hat\psi}{}_{[M}^i
\hat\gamma_{NP]}^{\vphantom{i}}\hat\epsilon_i^{\vphantom{i}}+\ft1{2\sqrt{5}}
\bar{\hat\lambda}{}^i
\hat\gamma_{MNP}^{\vphantom{i}}\hat\epsilon_i^{\vphantom{i}}],\nonumber\\
\delta\hat A_{M\,i}{}^j&=&\ft{i}{\sqrt{2}}e^{-\fft1{\sqrt{10}}\hat\phi}
[(\bar{\hat\psi}{}_M^j\hat\epsilon_i^{\vphantom{i}}-\ft1{\sqrt{5}}
\bar{\hat\lambda}{}^j
\hat\gamma_M^{\vphantom{i}}\hat\epsilon_i^{\vphantom{i}})-\ft12\delta_i{}^j
(\bar{\hat\psi}{}_M^j\hat\epsilon_k^{\vphantom{i}}-\ft1{\sqrt{5}}
\bar{\hat\lambda}{}^k
\hat\gamma_M^{\vphantom{i}}\hat\epsilon_k^{\vphantom{i}})],
\label{eq:7bsusy}
\end{eqnarray}
for the bosons.  Here, the supersymmetry transformation parameter $\hat
\epsilon_i$ is normalized according to
\begin{equation}
[\delta_1,\delta_2]\Xi = \ft14(\bar{\hat\epsilon}{}_2^i\hat\gamma^M
\hat\epsilon_{1\,i})\partial_M\Xi+(\hbox{general coordinate})+
(\hbox{local Lorentz})+(\hbox{gauge})\,,
\label{eq:susydef}
\end{equation}
where $\Xi$ represents any of the fields in the theory.

The $D=7$ spinors are symplectic-Majorana, with
$i,j=1,2$ being an $Sp(1)\equiv SU(2)$ index.  We take a convenient basis
where all $D=7$ Dirac matrices are antisymmetric, obeying
$\{\gamma^A,\gamma^B\}=2\eta^{AB}$.  The Majorana condition is
simply $\bar{\hat\lambda}{}^i=\epsilon^{ij}\hat\lambda_j^T$, and the
Majorana flip relation reads
\begin{equation}
\bar{\hat\chi}{}^i\gamma_{M_1M_2\cdots
M_n}^{\vphantom{i}}\hat\psi_i=(-)^n\bar{\hat\psi}{}^i\gamma_{M_nM_{n-1}\cdots
M_1}^{\vphantom{i}}
\hat\chi_i
\label{eq:mflip}
\end{equation}
(the triplet combination picks up an additional sign).  This spinor
convention is most convenient for reduction to $D=6$, as the
$D=7$ symplectic-Majorana spinors reduce trivially to their
six-dimensional counterparts.  Furthermore, an additional $D=6$ Weyl
condition may be imposed consistent with this Majorana condition, as
will be seen below.

   To reduce the fermions, we first examine the Killing spinors of the
domain wall background, given by (\ref{7to6ans}) with $\varphi$ and
$A_{(2)}$ set to zero.  Inserting this solution into (\ref{eq:7fsusy}),
we find
\begin{eqnarray}
\delta\hat\lambda_i&=&\ft1{\sqrt{10}}gW^{\fft15}(1+\gamma^7)\hat\epsilon_i,
\nonumber\\
\delta\hat\psi_{y\,i}&=&-\ft1{5\sqrt{2}}gW^{-1}\gamma^7(1-10g^{-1}
W\gamma^7\partial_y)
\hat\epsilon_i,\nonumber\\
\delta\hat\psi_{\mu\,i}&=&-\ft1{5\sqrt{2}}g\gamma_\mu(1+\gamma^7)
\hat\epsilon_i,
\end{eqnarray}
which leads to a half-BPS solution with Killing spinors given by
$\hat\epsilon_i=W^{-\fft1{10}}\epsilon_{0\,i}^{(-)}$.  Here,
$\epsilon_0^{(-)}$ is a constant six-dimensional symplectic-Majorana-Weyl
spinor with the chiral components defined by
\begin{equation}
\epsilon^{(\pm)}=P^{(\pm)}\epsilon\equiv\ft12(1\pm\gamma^7)\epsilon.
\end{equation}
Hence $\gamma^7$, the Dirac matrix in the $y$ direction, provides the
chirality operator on the brane.  This generation of chirality from a
non-chiral theory is a novel feature of this class of braneworld
reductions, and was previously investigated in
Refs.~\cite{bob1,bob2,dufliusab}.  Note that, were there
to be a modulus sign in $W$, the projection would instead be
$P^{(\pm)}=\fft12(1\pm\gamma^7{\rm sgn}\, y)$.  This provides an
obstruction to having globally well-defined Killing spinors, unless the
gauge coupling constant $g$ changes sign as well, thus compensating for
the sign change in $\partial_y W$ \cite{Bergshoeff:2000zn,Duff:2000az}.

Using the Killing spinors of the background as a guideline, it is then
straightforward to reduce the $D=7$ fermions.  There is one important
feature of the consistent reduction that needs mention, however.  Given
that the bosonic sector is that of a $(1,0)$ supergravity multiplet
coupled to a $(1,0)$ tensor multiplet, one must identify two chiral
spinors, $\psi_{\mu\,i}^{(-)}$ and $\lambda_i^{(+)}$, in six dimensions.
However a straightforward reduction of $\hat\psi_{M\,i}$ would suggest
the presence of an additional unwanted spin-1/2 field $\psi_{y\,i}^{(+)}$.
The resolution of this puzzle is that both $\hat\lambda$ and
$\psi_{y\,i}$ transform identically (up to factors), given the bosonic
ansatz (\ref{7to6ans}).  Thus they may be consistently set equal to one
another.  As a result, we obtain the reduction ansatz on the fermions
\begin{eqnarray}
\hat\psi_{\mu\,i}&=&W^{-\fft1{10}}e^{\fft12\alpha\varphi}[\psi_{\mu\,i}^{(-)}
+\ft1{10}\gamma_\mu\lambda_i^{(+)}]\,,\nonumber\\
\hat\psi_{y\,i}&=&-\ft25 W^{-\fft{11}{10}}e^{-\fft92\alpha\varphi}\gamma^7
\lambda_i^{(+)}\,,\nonumber\\
\hat\lambda_i&=&\ft2{\sqrt{5}}W^{\fft1{10}}e^{-\fft12\alpha\varphi}
\lambda_i^{(+)}\,,\nonumber\\
\hat\epsilon_i&=&W^{-\fft1{10}}e^{\fft12\alpha\varphi}\epsilon_i^{(-)}\,.
\label{eq:d76fred}
\end{eqnarray}
Although $\gamma^7$ has a definite eigenvalue when acting on definite
chirality spinors, we nevertheless retain it here and in the equations
below to avoid ambiguity in our choice of sign conventions.

Substitution of this ansatz into the $D=7$ fermion transformations,
(\ref{eq:7fsusy}), yields the $D=6$ transformations
\begin{eqnarray}
\delta\psi_{\mu\,i}^{(-)}&=&[\nabla_\mu-\ft1{48}e^{-\fft1{\sqrt{2}}\varphi}
F_{\nu\rho\sigma}\gamma^{\nu\rho\sigma}\gamma_\mu\gamma^7]\epsilon_i^{(-)}
\,,\nn\\
\delta\lambda_i^{(+)}&=&[-\ft1{2\sqrt{2}}\gamma^\mu\partial_\mu\varphi
+\ft1{24}e^{-\fft1{\sqrt{2}}\varphi}F_{\mu\nu\rho}\gamma^{\mu\nu\rho}
\gamma^7]\epsilon_i^{(-)}\,,
\label{eq:d76fsusy}
\end{eqnarray}
while substitution into the bosonic transformations, (\ref{eq:7bsusy}),
yields
\begin{eqnarray}
\delta\varphi&=&-\ft1{2\sqrt{2}}\bar\epsilon^{i\,(-)}\lambda_i^{(+)}\,,
\nonumber\\
\delta e_\mu^\alpha&=&\ft14\bar\epsilon^{i\,(-)}\gamma^\alpha
\psi_{\mu\,i}^{(-)}\,,\nonumber\\
\delta B_{\mu\nu}&=&\ft12e^{\fft1{\sqrt{2}}\varphi}(\bar\epsilon^{i\,(-)}
\gamma_{[\mu}^{\vphantom{()}}\gamma^7\psi_{\nu]\,i}^{(-)}
+\ft12\bar\epsilon^{i\,(-)}
\gamma_{\mu\nu}^{\vphantom{()}}\gamma^7\lambda_i^{(+)})\,.
\label{eq:d76bsusy}
\end{eqnarray}
Furthermore, consistency of the bosonic ansatz, (\ref{7to6ans}), is
maintained under supersymmetry, as we have verified that fields initially
set to zero remain so under their variations.  We see that these
transformations are simply those of ungauged ${\cal N}=(1,0)$ supergravity%
\footnote{One also sees that these transformations match the
appropriate truncation of the ${\cal N}=(1,1)$ transformations given below in
Eqs.~(\ref{eq:d6fsusy}) and (\ref{eq:d6bsusy}).}
in six dimensions \cite{Nishino:dc}
with supergravity multiplet
$(e_\mu^\alpha,\psi_{\mu\,i}^{(-)},B_{\mu\nu}^{(+)})$ and tensor
multiplet $(B_{\mu\nu}^{(-)},\lambda_i^{(+)},\varphi)$.

Additionally, consistent reduction of the fermion equations of motion
following from (\ref{eq:d7flag}) results in six-dimensional equations of
motion that may be derived from the Lagrangian
\begin{eqnarray}
{\cal L}_6^{\rm fermion}\!\!&=&\!\!-\ft12\bar\psi_\mu^i\gamma^{\mu\nu\rho}
\nabla_\nu
\psi_{\rho\,i}-\ft12\bar\lambda^i\gamma^\mu\nabla_\mu\lambda_i
-\ft18[\ft16\bar\psi_\mu^i\gamma^{\mu\nu\alpha\beta\gamma}\gamma^7
\psi_{\nu\,i}-\bar\psi^{\alpha\,i}\gamma^\beta\gamma^7\psi^\gamma_i]
e^{-\fft1{\sqrt{2}}\varphi}F_{\alpha\beta\gamma}\nonumber\\
&&-\ft18[\ft13\bar\psi_\mu^i\gamma^{\mu\alpha\beta\gamma}\gamma^7\lambda_i
-\bar\psi^{\alpha\,i}\gamma^{\beta\gamma}\gamma^7\lambda_i]
e^{-\fft1{\sqrt{2}}\varphi}F_{\alpha\beta\gamma}
-\ft1{2\sqrt{2}}(\bar\psi^i_\mu\gamma^\alpha\gamma^\mu\lambda_i)
\partial_\alpha\varphi\nonumber\\
&&+\ft1{48}(\bar\lambda^i\gamma^{\alpha\beta\gamma}
\gamma^7\lambda_i)
e^{-\fft1{\sqrt{2}}\varphi}F_{\alpha\beta\gamma}.
\end{eqnarray}
Along with the bosonic Lagrangian, (\ref{d6lag-1}), this reproduces the
ungauged six-dimensional $(1,0)$ model \cite{Nishino:dc}, up to
four-fermion terms.

\subsection{Lifting the dyonic string}

    The six-dimensional supergravity with tensor matter multiplet that
we have obtained via brane-world reduction admits a BPS dyonic string
solution.  By reversing the steps of the brane-world reduction we can
therefore obtain a BPS solution in the seven-dimensional gauged
supergravity.  Of course, this can be further lifted to ten-dimensions,
since the seven-dimensional gauged supergravity arises from an $S^3$
Scherk-Schwarz reduction of type I supergravity.

   The dyonic string solution in six dimensions is given by
\bea
ds_6^2 &=& (H_e\, H_m)^{-\fft12} \,(-dt^2 + dx^2) + (H_e\, H_m)^{\ft12}\,
(dr^2 + r^2\, d\Omega_3^2)\,,\nn\\
e^{\sqrt2\, \varphi} &=& \fft{H_m}{H_e}\,,\qquad
H_e = 1+\fft{Q}{r^2}\,,\qquad H_m = 1+ \fft{P}{r^2}\,,\\\
F_\3 &=& 2 P\, \Omega_\3 - dt\wedge dx \wedge d H_e^{-1}\,,\nn
\eea
where $\Omega_\3$ is the volume form of the 3-sphere in the transverse
space.  Lifting to $D=7$, we therefore obtain the solution
\bea
d\hat s_7^2 \!\!&=&\!\! W^{-\fft25}\, \Big( H_e^{-\fft25}\, H_m^{-\fft35}\, 
(-dt^2 + dx^2) + H_e^{\fft35}\, H_m^{\fft25}\,(dr^2 + r^2\, 
d\Omega_3^2)\Big) + W^{-\fft{12}{5}}\, H_e^{-\fft25}\, 
H_m^{\fft25}\, dy^2\!,\nn\\
e^{-\fft1{\sqrt{10}}\, \hat\phi} \!\!&=&\!\! \Big(\fft{W\, H_e}{H_m}\Big)
^{\fft15}\,,\\
\hat F_\4 \!\!&=&\!\! W^{-2}\, ( 2 P\, \Omega_\3  - dt\wedge dx\wedge 
dH_e^{-1})\wedge dy\,.\nn
\eea

   This seven-dimensional solution can be interpreted as the
intersection of a membrane and a string, living in the world-volume of
a 5-brane (domain wall). The corresponding harmonic functions are
$H_e$, $H_m$ and $W$ respectively.  In the standard Randall-Sundrum I
scenario, there are two distinct domain walls, with a separation $L$
in the vacuum state.  Turning on the radion mode will bend the walls
locally, giving them a space-time position dependent separation.  In
this particular BPS solution corresponding to the six-dimensional
dyonic string, the separation length is given by
\be 
L \sim \Big(\fft{H_m}{H_e}\Big)^{\fft15}= \Big(\fft{r^2 +
P}{r^2+Q}\Big)^{\fft15}\,.  
\ee
In the case $P=Q$, corresponding to the self-dual string, the radion
mode decouples and the separation is a constant.  For $P>Q$, the two
domain walls are convex, being closest at large $r$, with their
greatest separation occurring at $r=0$.  In the limit
$Q\longrightarrow 0$, which is a purely magnetic string, the
separation at $r=0$ becomes infinite.  Conversely, if $P<Q$ the domain
walls are concave, being closest at $r=0$.  The separation at $r=0$
becomes zero in the limit of a purely electric string, when
$P\longrightarrow 0$.  This demonstrates that a BPS configuration can
connect the visible world and the hidden brane, in the Randall-Sundrum
I scenario.

   It is worth remarking that a scalar potential characterised by 
$\Delta=-2$, such as we have here, is capable of trapping gravity
in a Randall-Sundrum II model \cite{youm1,youm2,cvlupors}.

\section{Further Supersymmetric Examples}\label{furthersusysec}

\subsection{$D=6$ reduced to $D=5$, ${\cal N}=2$ with vector multiplet}

    Our starting point is the six-dimensional gauged ${\cal N}=(1,1)$
supergravity.  The bosonic fields in this theory comprise the metric,
a dilaton $\phi$, a 2-form potential $A_\2$, and a 1-form potential
$B_\1$, together with the gauge potentials $A_\1^i$ of $SU(2)$
Yang-Mills.  The theory can be obtained from a consistent 
local $S^4$ reduction of massive type IIA supergravity \cite{6from10}.
The bosonic Lagrangian \cite{romans6}, converted to the
language of differential forms, is \cite{6from10}
\bea 
\hat {\cal L}_6 &=& \hat R\, {{\hat *}\oneone} -
\ft12 {{\hat *}d\hat\phi}\wedge
d\hat \phi - \ft29 g_2^2\, X^{-6} +\ft83 g_1\, g_2\, X^{-2} + 2 g_1^2\,
X^2\, {{\hat *}\oneone}\nn\\ 
&&-\ft12 X^4\, {{\hat *}\hat F_\3\wedge \hat F_\3}
-\ft12 X^{-2}\, \Big( {{\hat *}\hat G_\2}\wedge \hat G_\2 
+ {{\hat *}\hat F_\2^i}\wedge
\hat F_\2^i \Big) \label{d6lag}\\ 
&& - \hat A_\2\wedge(\ft12 d\hat B_\1\wedge d\hat B_\1
+\ft13 g_2\, \hat A_\2\wedge d\hat B_\1 
+\ft2{27} g_2^2\, \hat A_\2\wedge \hat A_\2 +\ft12
\hat F_\2^i\wedge \hat F_\2^i)\,,\nn 
\eea
where $X\equiv e^{-\hat \phi/(2\sqrt2)}$, $\hat F_\3= d\hat A_\2$,
$\hat G_\2= d\hat B_\1 + \ft23g_2\, \hat A_\2$, $\hat F_\2^i = d\hat
A_\1^i + \ft12 g_1\, \ep_{ijk} \hat A_\1^j\wedge \hat A_\1^k$, and
here ${\hat *}$ denotes the six-dimensional Hodge dual.  (We have
rescaled fields and coupling constants relative to the expression in
\cite{6from10}, to make explicit the gauge coupling $g_1$ and the mass
parameter $g_2$.)

In addition, the fermions comprise a symplectic-Majorana gravitino
$\hat\psi_{M\,i}$ and spinor $\hat\lambda_i$.  The $(1,1)$ supersymmetry
transformations are
\begin{eqnarray}
\delta\hat\psi_{M\,i}&=&[\hat D_M-\ft1{48}X^2\hat F_{ABC}\hat\gamma^{ABC}
\hat\gamma_M\gamma^7-\ft1{4\sqrt{2}}(g_1 X+\ft13g_2X^{-3})\hat\gamma_M
]\hat\epsilon_i\nn\\
&&\qquad
-\ft1{16\sqrt{2}}(\hat\gamma_M{}^{AB}-6\delta_M^A\hat\gamma^B)X^{-1}
(\hat G_{AB}\delta_i{}^j+i\gamma^7\hat F_{AB\,i}{}^j)\gamma^7\hat\epsilon_j
\,,\nn\\
\delta\hat\lambda_i&=&[-\ft1{2\sqrt{2}}\hat\gamma^M\partial_M\hat\phi
+\ft1{24}X^2\hat F_{MNP}\hat\gamma^{MNP}\gamma^7+\ft1{2\sqrt{2}}
(g_1X-g_2X^{-3})]\hat\epsilon_i\nn\\
&&\qquad+\ft1{8\sqrt{2}}X^{-1}(\hat G_{MN}\delta_i{}^j
+i\gamma^7\hat F_{MN\,i}{}^j)\hat\gamma^{MN}\gamma^7\hat\epsilon_j\,,
\label{eq:d6fsusy}
\end{eqnarray}
for the fermions, and
\begin{eqnarray}
\delta\hat e_M^A&=&\ft14\bar{\hat\epsilon}{}^i\gamma^A\hat\psi_{M\,i}\,,\nn\\
\delta\hat\phi&=&-\ft1{2\sqrt{2}}\bar{\hat\epsilon}{}^i\hat\lambda_i\,,\nn\\
\delta\hat B_M&=&\ft1{2\sqrt{2}}X(\bar{\hat\epsilon}{}^i\gamma^7
\hat\psi_{M\,i}-\ft12\bar{\hat\epsilon}{}^i\hat\gamma_M\gamma^7
\hat\lambda_i)\,,\nn\\
\delta\hat A_{MN}&=&\ft12X^{-2}(\bar{\hat\epsilon}{}^i\hat\gamma_{[M}
\gamma^7\hat\psi_{N]\,i}+\ft12\bar{\hat\epsilon}{}^i\hat\gamma_{MN}
\gamma^7\hat\lambda_i\,,\nn\\
\delta\hat A_{M\,i}{}^j&=&\ft{i}{\sqrt{2}}X[(\bar{\hat\epsilon}{}^j
\hat\psi_{M\,i}+\ft12\bar{\hat\epsilon}{}^j\hat\gamma_M\lambda_i)
-\ft12\delta_i{}^j(\bar{\hat\epsilon}{}^k
\hat\psi_{M\,k}+\ft12\bar{\hat\epsilon}{}^k\hat\gamma_M\lambda_k)]
\label{eq:d6bsusy}
\end{eqnarray}
for the bosons.  Our convention for symplectic-Majorana spinors
parallels that given above for the $D=7$ case.   In particular,
the Majorana flip relation is identical to (\ref{eq:mflip}).
Additionally, we follow the same normalization as (\ref{eq:susydef}).
Here, as in seven dimensions, the gauge covariant derivative acting
on an $Sp(1)$ spinor is defined as
$\hat D_M\hat\epsilon_i=\hat\nabla_M\hat\epsilon_i+\fft{i}2g_1\hat
A_{M\,i}{}^j,$
where $\hat A_{M\,i}{}^j\equiv \hat A_M^k(-\sigma^k)_i{}^j$, so that
$\hat F_{MN\,i}{}^j=\partial_M\hat A_{N\,i}{}^j+\fft{i}2g
\hat A_{M\,i}{}^k\hat A_{N\,k}{}^j-({\scriptstyle M}\leftrightarrow
{\scriptstyle N})$.  

    For our brane-world reduction, we shall take the mass parameter
$g_2$ to zero, and relabel $g_1$ as $g$, giving
\bea 
\hat {\cal L}_6 &=& \hat R\, {{\hat *}\oneone} 
-\ft12 {{\hat *}d\hat\phi}\wedge d\hat \phi + 2 g^2\,
X^2\, {{\hat *}\oneone}\nn\\ 
&&-\ft12 X^4\, {{\hat *} \hat F_\3\wedge \hat F_\3}
-\ft12 X^{-2}\, \Big( {{\hat *}\hat G_\2}\wedge \hat G_\2 
+ {{\hat *}\hat F_\2^i}\wedge
\hat F_\2^i \Big) \label{d6lag-2}\\ 
&& - \hat A_\2\wedge(\ft12 d\hat B_\1\wedge d\hat B_\1
+\ft12 \hat F_\2^i\wedge \hat F_\2^i)\,,\nn 
\eea
with $\hat G_\2= d\hat B_\1$, $\hat F_\2^i = d\hat A_\1^i + 
\ft12 g\, \ep_{ijk} \hat A_\1^j\wedge \hat A_\1^k$.  
From section 2, we are then led to make the reduction
ansatz
\bea
d\hat s_6^2 &=& W^{-\fft12}\, e^{2\alpha\,\varphi}\, ds_5^2 + W^{-\fft52}\, 
e^{-6\alpha\,\varphi}\, dy^2\,,\nn\\
e^{-\fft1{\sqrt2}\, \hat\phi} &=& W^{\fft12}\, e^{6\alpha\, 
\varphi}\,,\nn\\
\hat A_\2 &=& W^{-2}\, A_1\wedge dy\,,\qquad 
\hat B_1 = B_1\,,\qquad \hat A_\1^i=0\,,\label{6to5ans}
\eea
with $\alpha= -1/(4\sqrt6)$ and $W=1+\sqrt{2}gy$.  Substituting
into the equations of
motion that follow from (\ref{d6lag-2}), we find consistent
five-dimensional equations of motion that can be derived from the
Lagrangian
\be
{\cal L}_5 = R\, {*\oneone} - \ft12 {*d\varphi}\wedge d\varphi 
-\ft12 e^{-\fft{4}{\sqrt6}\,\varphi}\, {*F_\2}\wedge F_\2 -
\ft12 e^{\fft{2}{\sqrt6}\, \varphi}\, {*G_\2}\wedge G_\2 -\ft12
A_\1\wedge dB_\1\wedge dB_\1\,,
\ee
where $F_\2=dA_\1$ and $G_2=dB_\1$.  This is the bosonic sector of
ungauged ${\cal N}=2$ five-dimensional supergravity, coupled to a vector
multiplet whose bosonic fields are $\varphi$ and $B_1$.

This identification of the resulting theory with five-dimensional ${\cal N}=2$
supergravity may be confirmed by reducing the fermionic sector.
Applying the procedure outlined in section~\ref{7to6red} to the present
fermions, we find the appropriate reduction to be
\begin{eqnarray}
\hat\psi_{\mu\,i}&=&W^{-\fft13}e^{\fft12\alpha\varphi}[\psi_\mu^{(-)}
+\ft1{2\sqrt{6}}\gamma_\mu\lambda^{(+)}]\,,\nn\\
\hat\psi_{y\,i}&=&-\ft3{2\sqrt{6}}W^{-\fft98}e^{-\fft72\alpha\varphi}
\gamma^6\lambda^{(+)}\,,\nn\\
\hat\lambda_i&=&\ft3{\sqrt{6}}W^{\fft18}e^{-\fft12\alpha\varphi}
\lambda^{(+)}\,,\nn\\
\hat\epsilon_i&=&W^{-\fft18}e^{\fft12\alpha\varphi}
\epsilon^{(-)}\,,
\end{eqnarray}
where the `chirality' is determined by the projection
$P^{(\pm)}=\fft12(1\pm\gamma^6)$.  For the moment, we retain the
six-dimensional form of the spinors and Dirac matrices.  The reduction
of the fermion transformations, (\ref{eq:d6fsusy}) yields
\begin{eqnarray}
\delta\psi_{\mu\,i}^{(-)}&=&[\nabla_\mu-\ft1{24}(\gamma_\mu{}^{\nu\lambda}
-4\delta_\mu^\nu\gamma^\lambda)(\sqrt{2}e^{\fft1{\sqrt{6}}\varphi}
G_{\nu\lambda}-e^{-\fft2{\sqrt{6}}\varphi}F_{\nu\lambda}\gamma^6)\gamma^7]
\epsilon_i\,,\nn\\
\delta\lambda_i^{(+)}&=&[-\ft14\gamma^\mu\partial_\mu\varphi
+\ft1{8\sqrt{3}}(e^{\fft1{\sqrt{6}}\varphi}G_{\mu\nu}+\sqrt{2}
e^{-\fft2{\sqrt{6}}\varphi}F_{\mu\nu}\gamma^6)\gamma^{\mu\nu}\gamma^7]
\epsilon_i\,,
\label{eq:d6fsred}
\end{eqnarray}
while the reduction of the boson transformations, (\ref{eq:d6bsusy})
yields
\begin{eqnarray}
\delta e_\mu^\alpha&=&\ft14\bar\epsilon^i\gamma^\alpha\psi_{\mu\,i}\,,\nn\\
\delta\varphi&=&-\ft12\bar\epsilon^i\lambda_i\,,\nn\\
\delta A_\mu&=&\ft12e^{\fft2{\sqrt{6}}\varphi}(\ft2{\sqrt{6}}\bar\epsilon^i
\gamma_\mu\gamma^6\gamma^7\lambda_i-\ft12\bar\epsilon^i\gamma^6\gamma^7
\psi_{\mu\,i})\,,\nn\\
\delta B_\mu&=&-\ft1{\sqrt{2}}e^{-\fft1{\sqrt{6}}\varphi}
(\ft1{\sqrt{6}}\bar\epsilon^i\gamma_\mu\gamma^7\lambda_i
-\ft12\bar\epsilon^i\gamma^7\psi_{\mu\,i})\,.
\label{eq:d6bsred}
\end{eqnarray}
Again, we find that this reduction is consistent.

Finally, to make connection to the $D=5$, ${\cal N}=2$ supergravity, we rewrite
the fermions in terms of natural five-dimensional spinors.  To do so, we
first note that the expressions in (\ref{eq:d6fsred}) and
(\ref{eq:d6bsred}) are trivial under $Sp(1)$, and hence the
symplectic-Majorana spinors may be combined into ordinary six-dimensional
Dirac spinors.  Then we may choose a decomposition of the Dirac matrices
as, {\it e.g.}
\begin{eqnarray}
\gamma^\mu&=&\tilde\gamma^\mu\times\sigma^1,\qquad \mu=0,1,\ldots,4\,,\nn\\
\gamma^6&=&1\times\sigma^3\,,\nn\\
\gamma^7\equiv\gamma^{01234}\gamma^6&=&1\times\sigma^2\,.
\label{eq:d75gam}
\end{eqnarray}
Here, $\tilde\gamma^\mu$ are a set of $4\times4$ Dirac matrices for
$D=5$.  Note that `chiral' spinors under $P^{(\pm)}$ may be written
as
\begin{equation}
\chi^{(+)}=\pmatrix{\chi\cr0}\,,\qquad\chi^{(-)}=\pmatrix{0\cr\chi}\,.
\end{equation}
Each five-dimensional Dirac matrix flips $P^{(\pm)}$ chirality, and
furthermore, we have $\gamma^7\chi^{(\pm)}=\pm i\chi^{(\mp)}$.  As a
result, Eqs.~(\ref{eq:d6fsred}) and (\ref{eq:d6bsred}) take on the
five-dimensional form
\begin{eqnarray}
\delta\psi_\mu&=&[\nabla_\mu+\ft{i}{24}(\gamma_\mu{}^{\nu\lambda}
-4\delta_\mu^\nu\gamma^\lambda)(\sqrt{2}e^{\fft1{\sqrt{6}}\varphi}
G_{\nu\lambda}-e^{-\fft2{\sqrt{6}}\varphi}F_{\nu\lambda})]\epsilon\,,\nn\\
\delta\lambda&=&[-\ft14\gamma^\mu\partial_\mu\varphi
-\ft{i}{8\sqrt{3}}(e^{\fft1{\sqrt{6}}\varphi}G_{\mu\nu}+\sqrt{2}
e^{-\fft2{\sqrt{6}}\varphi}F_{\mu\nu})\gamma^{\mu\nu}]\epsilon\,,\nn\\
\delta e_\mu^\alpha&=&\ft14\bar\epsilon\gamma^\alpha\psi_\mu\,,\nn\\
\delta\varphi&=&-\ft12\bar\epsilon\lambda\,,\nn\\
\delta A_\mu&=&-\ft12e^{\fft2{\sqrt{6}}\varphi}(\ft{2i}{\sqrt{6}}\bar\epsilon
\gamma_\mu\lambda-\ft{i}2\bar\epsilon\psi_\mu)\,,\nn\\
\delta B_\mu&=&\ft1{\sqrt{2}}e^{-\fft1{\sqrt{6}}\varphi}
(-\ft{i}{\sqrt{6}}\bar\epsilon^i\gamma_\mu\lambda
-\ft{i}2\bar\epsilon\psi_\mu)\,,
\end{eqnarray}
where we have now dropped the tildes on the five-dimensional Dirac
matrices.  After transforming to a Dirac normalization and taking
$\lambda\to-i\lambda$, these transformations agree with those of
${\cal N}=2$ supergravity coupled to a single vector multiplet,
as given below in (\ref{eq:5susy}).

   One can construct BPS solutions (strings or black holes) in five
dimensions, and then lift them back to $D=6$. Again, the bending of
the domain walls will be convex or concave according to the relative
sizes of the two charges carried by $F_\2$ and $G_\2$.

\subsection{$D=5$ reduced to $D=4$, ${\cal N}=1$ with a chiral multiplet}
\label{5to4susy}

In this example, we shall take as the starting point the five-dimensional
${\cal N}=2$ gauged supergravity coupled to a vector multiplet.  In general,
this ${\cal N}=2$ theory is described by very special geometry
\cite{Gunaydin:1983bi,Gunaydin:1984ak}, and we begin
with a brief outline of some relevant facts.  For the coupling of
supergravity to $n$ vector multiplets, in addition to the graviton $\hat
g_{MN}$ and gravitino $\hat\psi_M$, one introduces $n+1$ vector
potentials $\hat A_{(1)}^I$, as well as $n$ scalars $\hat\phi^i$ and
gauginos $\hat\lambda^i$.  Of the $n+1$ vectors, the ${\cal N}=2$ graviphoton
is given by the linear combination $\hat{\cal A}_{(1)}=V_IA^I_{(1)}$,
where the $V_I$ are a set of constants related to the gauging.

The bosonic Lagrangian takes the form
\begin{equation}
\hat {\cal L}_5 = \hat R\, {\hat *\oneone} -
{\cal G}_{ij} {\hat *d\hat\phi^i}\wedge d\hat\phi^j
+ 2g^2\,V \, {\hat *\oneone}
-G_{IJ}\, {\hat *\hat F_\2^I}\wedge \hat F_\2^J
-\ft16 C_{IJK} \hat A_\1^I\wedge d\hat B_\1^J\wedge d\hat B_\1^K\,,
\end{equation}
where the constants $C_{IJK}$ specify a homogeneous cubic polynomial
\begin{equation}
{\cal V}=\fft16C_{IJK}X^IX^JX^K\,.
\end{equation}
Here, the $n+1$ quantities $X^I$ are functions of the $n$ scalar fields
$\phi^i$, and are required to satisfy the condition ${\cal V}=1$.
The quantities $G_{IJ}$ and ${\cal G}_{ij}$ in the Lagrangian are given by
\begin{eqnarray}
G_{IJ}&=&-\ft12\partial_I\partial_J\log{\cal V}\Big|_{{\cal V}=1}\,,\nn\\
{\cal G}_{ij}&=&\partial_iX^I\partial_jX^JG_{IJ}\Big|_{{\cal V}=1}\,,
\end{eqnarray}
and the potential, which arises from the gauging has the form
\begin{eqnarray}
V=V_IV_J(6X^IX^J-\ft92{\cal G}^{ij}\partial_iX^I\partial_jX^J)\,.
\end{eqnarray}
For a more complete treatment, see
\cite{Gunaydin:1983bi,Gunaydin:1984ak}.

In terms of a single five-dimensional Dirac spinor, the ${\cal N}=2$
supersymmetry transformations are
\begin{eqnarray}
\delta\hat\psi_M&=&[\hat D_M+\ft{i}8(\hat\gamma_M{}^{NP}-4\delta_M^N
\hat\gamma^P)X_I\hat F_{NP}^I+\ft12g\hat\gamma_MX^IV_I]\hat\epsilon\,,\nn\\
\delta\hat\lambda_i&=&\partial_iX^I[-\ft14G_{IJ}\hat\gamma^{MN}
\hat F_{MN}^I+\ft{3i}4 \hat\gamma^M\partial_MX_I+\ft{3i}2gV_I]\hat\epsilon
\,,\nn\\
\delta\hat e_M^A&=&\ft12\bar{\hat\epsilon}\gamma^A\hat\psi_M\,,\nn\\
\delta\hat A_M^I&=&\ft12{\cal G}^{ij}\partial_iX^I\bar{\hat\epsilon}
\hat\gamma_M\hat\lambda-\ft{i}2X^I\bar{\hat\epsilon}\hat\psi_M\,,\nn\\
\delta\hat\phi^i&=&\ft{i}2{\cal G}^{ij}\bar{\hat\epsilon}\hat\lambda_j\,.
\end{eqnarray}
The supersymmetry transformation parameter $\hat\epsilon$ is normalized
according to
\begin{equation}
[\delta_1,\delta_2]\Xi=\ft12(\bar{\hat\epsilon}{}_2\hat\gamma^M
\hat\epsilon_1)\partial_M\Xi+\cdots\,,
\end{equation}
and the gauge covariant derivative acting on a charged spinor is given by
\begin{equation}
\hat D_M=\hat\nabla_M-\ft{3i}2g\hat{\cal A}_M=\hat\nabla_M-\ft{3i}2g
V_I\hat A^I_M\,.
\end{equation}

With these preliminaries out of the way, we now focus on the model on
hand, namely gauged supergravity coupled to a single vector multiplet
({\it i.e.}~$n=1$).  This model is obtained by taking
$C_{112}=C_{121}=C_{211}=1$, and by specifying the gauging according to
\begin{equation}
gV_1=\fft{\sqrt{2}}6g_1,\qquad gV_2=\fft16g_2\,.
\end{equation}
A convenient scalar parametrization preserving ${\cal V}=1$ is then
\begin{equation}
X^1=\sqrt{2}e^{-\fft1{\sqrt{6}}\hat\phi},\qquad
X^2=e^{\fft2{\sqrt{6}}\hat\phi}\,.
\end{equation}
We furthermore define $\hat F_\2^1=\hat G_\2=d\hat
B_\1$, and $\hat F_\2^2=\hat F_\2=d\hat A_\1$.  The Lagrangian of the
bosonic sector is then given by
\bea
\hat {\cal L}_5 &=& \hat R\, {\hat *\oneone} -
\ft12 {\hat *d\hat\phi}\wedge d\hat\phi
+ (2g_1g_2\, e^{\fft{1}{\sqrt6}\,\hat\phi} +
          g_1^2\,e^{-\fft{2}{\sqrt6}\,\hat\phi})\, {\hat *\oneone}\nn\\
&&-\ft12 e^{-\fft{4}{\sqrt6}\,\hat\phi}\, {\hat *\hat F_\2}\wedge \hat F_\2 -
\ft12 e^{\fft{2}{\sqrt6}\, \hat\phi}\, {\hat *\hat G_\2}\wedge
\hat G_\2 -\ft12 \hat A_\1\wedge d\hat B_\1\wedge d\hat B_\1\,,\label{d5lag}
\eea
with supersymmetry transformations
\begin{eqnarray}
\delta\hat\psi_M&=&[\hat D_M+\ft{i}{24}(\hat\gamma_M{}^{NP}-4\delta_M^N
\hat\gamma^P)(\sqrt{2}e^{\fft1{\sqrt{6}}\hat\phi}\hat G_{NP}+
e^{-\fft2{\sqrt{6}}\hat\phi}\hat F_{NP})\nn\\
&&\kern6cm+\ft1{12}\hat\gamma_M
(2g_1e^{-\fft1{\sqrt{6}}\hat\phi}+g_2e^{\fft2{\sqrt{6}}\hat\phi})]
\hat\epsilon\,,\nn\\
\delta\hat\lambda&=&[-\ft{i}4\hat\gamma^M\partial_M\hat\phi+\ft1{8\sqrt{3}}
\hat\gamma^{MN}(e^{\fft1{\sqrt{6}}\hat\phi}\hat G_{MN}
-\sqrt{2}e^{-\fft2{\sqrt{6}}\hat\phi}\hat F_{MN})\nn\\
&&\kern6cm-\ft{i}{2\sqrt{6}}
(g_1e^{-\fft1{\sqrt{6}}\hat\phi}-g_2e^{\fft2{\sqrt{6}}\hat\phi})
]\hat\epsilon\,,\nn\\
\delta\hat e_M^A&=&\ft12\bar{\hat\epsilon}\gamma^A\hat\psi_M\,,\nn\\
\delta\hat\phi&=&i\bar{\hat\epsilon}\hat\lambda\,,\nn\\
\delta\hat A_M&=&e^{\fft2{\sqrt{6}}\hat\phi}(\ft2{\sqrt{6}}
\bar{\hat\epsilon}\hat\gamma_M\hat\lambda-\ft{i}2\bar{\hat\epsilon}
\hat\psi_M)\,,\nn\\
\delta\hat B_M&=&\sqrt{2}e^{-\fft1{\sqrt{6}}\hat\phi}(-\ft1{\sqrt{6}}
\bar{\hat\epsilon}\hat\gamma_M\hat\lambda-\ft{i}2\bar{\hat\epsilon}
\hat\psi_M)\,.
\label{eq:5susy}
\end{eqnarray}
The explicit embedding in eleven dimensional supergravity of this
AdS$_5$ supergravity coupled to a vector multiplet was given in
\cite{10authors}.

        For our present purposes, we shall set $g_2=0$ and relabel $g_1$
as $g$.  The resulting theory has a domain wall as its vacuum
solution instead of AdS$_5$.  As shown in appendix \ref{s2red}, this
domain wall supergravity can be obtained from the $S^2$ reduction of
the seven-dimensional domain-wall supergravity (\ref{d7lag-2})
discussed in section~\ref{7to6red}.  From the results in section 2, we
are led to make the following Kaluza-Klein reduction ansatz:
\bea
d\hat s_5^2 &=& W^{-\fft23}\, e^{2\alpha\, \varphi}\, ds_6^2 +
W^{-\fft83}\, e^{-4\alpha\,\varphi}\, dy^2\,,\nn\\
e^{-\fft2{\sqrt 6}\, \hat\phi} &=& W^{\fft23}\, e^{4\alpha\, \varphi}\,,\nn\\
\hat A_\1 &=& \chi\, W^{-2}\, dy\,,\qquad \hat B_\1=0\,,\label{5to4ans}
\eea
where $\a=\ft16$ and $m^2=g^2$ (so that $W=1+gy$).  Substituting into
the equations of motion following from (\ref{d5lag}) (with $g_1=g$,
$g_2=0$), we obtain a consistent reduction of the bosonic fields to a
four-dimensional system whose equations of motion can be derived from
the Lagrangian
\be
{\cal L}_4 = R\, {*\oneone} - \ft12 {*d\varphi}\wedge d\varphi
-\ft12 e^{2\varphi}\, {*d\chi}\wedge d\chi\,.
\ee
This is precisely the bosonic sector of four-dimensional ${\cal N}=1$
supergravity coupled to a chiral scalar multiplet.

The fermionic reduction may similarly be obtained following the method
developed in section~\ref{7to6red}.  Defining here the four-dimensional
chirality
projection $P^{(\pm)}=\fft12(1\pm\gamma^5)$, the appropriate reduction
on the fermions is given by
\begin{eqnarray}
\hat\psi_\mu&=&W^{-\fft16}e^{\fft1{12}\varphi}[\psi_\mu^{(+)}
-\ft{i}6\gamma_\mu\lambda^{(-)}]\,,\nn\\
\hat\psi_y&=&\ft{i}3W^{-\fft76}e^{-\fft5{12}\varphi}\gamma^5\lambda^{(-)}
\,,\nn\\
\hat\lambda&=&\ft1{\sqrt{6}}W^{\fft16}e^{-\fft1{12}\varphi}\lambda^{(-)}
\,,\nn\\
\hat\epsilon&=&W^{-\fft16}e^{\fft1{12}\varphi}\epsilon^{(+)}\,.
\label{eq:d54fred}
\end{eqnarray}
The reduction of the five-dimensional supersymmetry transformations,
(\ref{eq:5susy}), yields
\begin{equation}
\delta\psi_\mu^{(+)}=[\nabla_\mu-\ft{i}4e^\varphi\partial_\mu\chi
\gamma^5]\epsilon^{(+)}\,,\qquad
\delta e_\mu^\alpha=\ft12\bar\epsilon^{(+)}\gamma^\alpha
\psi_\mu^{(+)}\,,
\end{equation}
for the four-dimensional supergravity multiplet, and
\begin{eqnarray}
\delta\lambda^{(-)}&=&\ft12\gamma^\mu[i\partial_\mu\varphi-e^\varphi
\partial_\mu\chi\gamma^5]\epsilon^{(+)}\,,\nn\\
\delta\varphi&=&-\ft{i}2\bar\epsilon^{(+)}\lambda^{(-)}\,,\nn\\
\delta\chi&=&\ft12e^{-\varphi}\bar\epsilon^{(+)}\gamma^5\lambda^{(-)}\,.
\end{eqnarray}
for the matter fields.  Again, we have verified that this is a
consistent reduction on the fermions.  This confirms our identification
of the reduced theory as ${\cal N}=1$ supergravity coupled to a chiral
multiplet.  Standard techniques may be used to rewrite the
four-dimensional Weyl spinors in terms of Majorana ones.

\section{Extended Bosonic Examples}
\label{sec:ebe}

   In this section, we show that it is possible to enlarge the bosonic
ans\"atze of the previous consistent reductions in sections
\ref{7to6red} and \ref{furthersusysec}, to obtain bosonic reductions
that yield larger numbers of fields from the same higher-dimensional
starting points.  Although these larger bosonic reductions are still
fully consistent, we find that they are no longer the bosonic sectors
of supergravities in the lower dimensions.  In other words, although
the bosonic sectors of the higher-dimensional supergravities admit
enlarged consistent reductions that retain more lower-dimensional
fields, it is not possible to make corresponding enlarged consistent
reductions in the fermionic sectors.  Nonetheless, the fact that the
bosonic sectors admit enlarged consistent reductions is of interest in
its own right.

\subsection{$D=7$ reduced with $SU(2)$ Yang-Mills in $D=6$}
\label{sec:7to6su2}

   In this enlarged consistent reduction, we begin with the same
seven-dimensional Lagrangian (\ref{d7lag-2}) that we used in section
\ref{7to6red}, but we now include the $SU(2)$ Yang-Mills fields as
well, which were previously set to zero in (\ref{7to6ans}).  From the
results in section 2, we see that a consistent reduction for the $\hat
A_\1^i$ potentials should be possible, with the reduction ansatz given
by
\be
\hat A_i= A_i\,.
\label{eq:su2red}
\ee
It is easy to verify that within the bosonic sector, the entire 
reduction, with ansatz given by (\ref{7to6ans}) except that 
$\hat A_\1^i =A_\1^i$ instead of $\hat A_\1^i=0$, is
consistent, and the resulting bosonic Lagrangian in $D=6$ is given by
\be
{\cal L}_6 = R\, {*\oneone} - \ft12 {*d\varphi}\wedge d\varphi 
- \ft12 e^{-\sqrt2\, \varphi}\, {*F_3}\wedge F_3
-\ft12 e^{\fft1{\sqrt2}\, \varphi}\, {*F_\2^i}\wedge F_\2^i 
+\ft12 F_\2^i\wedge F_\2^i\wedge A_\2\,,\label{d6lag-nonsusy}
\ee
where $F_\3=dA_\2$ and $F_\2^i=dA_\1^i + \ft12 g\, \epsilon_{ijk}\,
A_\1^i\wedge A_\1^k$.

   We have obtained a lower-dimensional theory that includes $SU(2)$
Yang-Mills fields, but where nevertheless there is no scalar
potential.  This result is rather surprising.  The $D=7$ gauged
supergravity itself can be obtained from ${\cal N}=1$, $D=10$
supergravity, which is ungauged, by reduction on $S^3$.  Thus by
combining a standard Scherk-Schwarz group manifold reduction with an
additional stage of brane-world reduction, we can obtain a theory with
non-abelian Yang-Mills fields coming from the geometry of the internal
space, and yet this lower-dimensional theory has no cosmological term.

   At first sight the Lagrangian (\ref{d6lag-nonsusy}) appears to be
precisely the bosonic sector of $D=6$ $(1,0)$ supergravity coupled to
a tensor multiplet together with an $SU(2)$ adjoint vector multiplet.
However, this is in fact not the case.  One way to see this is to note that,
while the lower-dimensional theory has no potential, the original $D=7$
gauged supergravity has gravitini charged under this same $SU(2)$ (which
is in fact the $Sp(1)$ symmetry of the symplectic Majorana spinors).
Dimensional reduction along the lines of (\ref{eq:su2red}) can never
remove the charge of the gravitini.  Thus, were the resulting theory to
be supersymmetric, one would end up with a $D=6$, ${\cal N}=(1,0)$ gauged
supergravity without a scalar potential.  The fact that there are no known
theories of this form immediately provides a hint that the reduction
cannot be supersymmetric.

     Of course, one may entertain the possibility that such a class of
gauged (or `partially gauged') supergravities might in fact exist.
Thus it is worth examining the supersymmetry of the extended reduction
in some detail.  Since the domain wall vacuum preserves ${\cal
N}=(1,0)$, we are still concerned with only a chiral supersymmetry,
parameterized by $\epsilon_i^{(-)}$.  However, now additional
chiralities show up in the fermion reduction.  Relaxing
(\ref{eq:d76fred}) to include both chiralities of the $D=6$ spinors
(but retaining the connection between $\hat\psi_{y\,i}$ and
$\hat\lambda_i$, which is still consistent), we obtain, in addition to
(\ref{eq:d76fsusy}) with $\nabla_\mu$ replaced by $D_\mu$ and
(\ref{eq:d76bsusy}), the transformations
\begin{eqnarray}
\delta\psi_{\mu\,i}^{(+)}&=&[\ft{i}{16}e^{\fft1{2\sqrt{2}}\varphi}
(\gamma_\mu{}^{\alpha\beta}-6\delta_\mu^\alpha\gamma^\beta)
F_{\alpha\beta\,i}{}^j]\epsilon_j^{(-)}\,,\nn\\
\delta\lambda_i^{(-)}&=&[-\ft{i}8e^{\fft1{2\sqrt{2}}\varphi}
F_{\mu\nu\,i}{}^j\gamma^{\mu\nu}]\epsilon_j^{(-)}\,,\\
\delta A_{\mu\,i}{}^j&=&\ft{i}2e^{-\fft1{2\sqrt{2}}\varphi}
[(\bar\psi_\mu^{j\,(+)}\epsilon_i^{(-)}-\ft12\bar\lambda^{j\,(-)}
\gamma_\mu\epsilon_i^{(-)})-\ft12\delta_i{}^j
(\bar\psi_\mu^{k\,(+)}\epsilon_k^{(-)}-\ft12\bar\lambda^{k\,(-)}
\gamma_\mu\epsilon_k^{(-)})]\,.\nn
\end{eqnarray}
At first, this is quite intriguing, as this suggests that the $SU(2)$
vectors are in fact grouped into a spin-$\fft32$ multiplet.
Essentially, since the gauge fields were superpartners of the gravitini
in $D=7$, they remain superpartners of spin-$\fft32$ matter in the
reduced $D=6$ theory.  However, this identification presupposes the
existence of an abelian vector which is lacking in the reduction, as
the spin-$\fft32$ multiplet consists of the fields $(\psi^{(+)}_{\mu\,i},
A_{\mu\,i}{}^j,A_\mu,\lambda_i^{(-)})$.

Another way to see that this reduction cannot be supersymmetric is to
note that the presence of $\psi_{\mu\,i}^{(+)}$ and $\lambda_i^{(-)}$
yields the $D=7$ transformations
\begin{eqnarray}
\delta\hat g_{\mu y}&=&\ft14W^{-\fft75}e^{\fft3{10\sqrt{2}}\varphi}
\bar\epsilon^{i\,(-)}\gamma^7(\psi_{\mu\,i}^{(+)}+\ft12\gamma_\mu
\lambda_i^{(-)})\,,\nn\\
\delta\hat A_{\mu\nu\rho}&=&-\ft34W^{-1}e^{\fft1{2\sqrt{2}}\varphi}
\bar\epsilon^{i\,(-)}(\gamma_{[\mu\nu}\psi_{\rho]\,i}^{(+)}-\ft16
\gamma_{\mu\nu\rho}\lambda_i^{(-)})\,.
\label{eq:icsusy}
\end{eqnarray}
So although both $\hat g_{\mu\,y}$ and $\hat A_{\mu\nu\rho}$ are set to
zero in the reduction ansatz (\ref{7to6ans}), this structure cannot be
maintained under supersymmetry transformations in the presence of these
additional fields.
Consistency of the reduction under supersymmetry then requires the
vanishing of $A_{\mu\,i}{}^j$, $\psi_{\mu\,i}^{(+)}$, and $\lambda_i^{(-)}$.  
Nevertheless, the form of (\ref{eq:icsusy}) suggests a possible further
generalization of the reduction ansatz to include a vector arising
from an appropriate linear combination of the dual of $\hat A_{\mu\nu\rho}$
and an off-diagonal metric component $\hat g_{\mu y}$.  This possibility
is currently under investigation.

\subsection{$D=6$ reduced with $SU(2)$ Yang-Mills in $D=5$}

    Analogously, the $SU(2)$ Yang-Mills fields in the $D=6$ Lagrangian
(\ref{d6lag-2}) can also be consistently reduced.  The complete
reduction ansatz is given by (\ref{6to5ans}), except that now $\hat
A_\1^i=A_\1^i$ instead of being set to zero.  The resulting Lagrangian
in $D=5$ is given by
\bea
{\cal L}_5 &=& R\, {*\oneone} - \ft12 {*d\varphi}\wedge d\varphi 
-\ft12 e^{-\fft{4}{\sqrt6}\,\varphi}\, {*F_\2}\wedge F_\2 -
\ft12 e^{\fft{2}{\sqrt6}\, \varphi}\, \Big({*G_\2}\wedge G_\2 +
{*F_\2^i}\wedge F_\2^i\Big)\nn\\
&&-\ft12
A_\1\wedge \Big(dB_\1\wedge dB_\1 + F_\2^i\wedge F_\2^i\Big)\,,
\eea
where $F_\2=dA_\1$, $G_\2=dB_\1$ and $F_\2^i=dA_\1^i + \ft12 g\,
\epsilon_{ijk}\, A_\1^j\wedge A_\1^k$.

\subsection{$D=5$ reduced with an additional vector}

     The vector $B_\1$ in section \ref{5to4susy} can also be included
in a consistent reduction, with the ansatz given by (\ref{5to4ans})
except that $\hat B_\1=B_\1$.  The resulting Lagrangian is now given
by
\be
{\cal L}_4 = R\, {*\oneone} - \ft12 {*d\varphi}\wedge d\varphi 
-\ft12 e^{2\varphi}\, {*d\chi}\wedge d\chi -\ft12 e^{-\varphi}\,
{*G_\2}\wedge G_\2 +\ft12 \chi\, G_\2\wedge G_\2\,,
\ee
where $G_\2=dB_\1$.

         At first sight, this is exactly the bosonic Lagrangian of
${\cal N}=1$ supergravity coupled to a vector and a chiral multiplet.
On the other hand, this runs into a similar difficulty with supersymmetry
as found above in section \ref{sec:7to6su2}.  Were supersymmetry to be
valid, somehow, one again runs into a problem identifying the
superpartner to the vector as either spin-$\fft12$ or spin-$\fft32$.  In
this case, the generalization of (\ref{eq:d54fred}) yields the additional
transformations
\begin{eqnarray}
\delta\psi_\mu^{(-)}&=&\ft{i}{8\sqrt{2}}e^{-\fft12\varphi}G_{\alpha\beta}
\gamma^{\alpha\beta} \gamma_\mu \epsilon^{(+)}\,,\nn\\
\delta\lambda^{(+)}&=&\ft1{8\sqrt{3}}e^{-\fft12\varphi}G_{\alpha\beta}
\gamma^{\alpha\beta}\epsilon^{(+)}\,,
\label{eq:d54icf}
\end{eqnarray}
which is suggestive of both spin-$\fft12$ and spin-$\fft32$
simultaneously.  Note here that the possibility of consistently
setting $\lambda^{(+)}$ to be proportional to $\gamma^\mu\psi_\mu^{(-)}$
will not work, as the latter is kinematically vanishing.  Similarly, one
finds by supersymmetry that both $\hat g_{\mu y}$ and $\hat A_\mu$
cannot be consistently set to zero unless both $\psi_\mu^{(-)}$ and
$\lambda^{(+)}$ are absent.  So we again conclude that the extend
reduction ansatz here is inconsistent with the inclusion of the
fermions.

Unlike for the extended $D=7$ to $D=6$ case, however, (which
lacks even the requisite bosonic fields for supersymmetry) here the
bosonic sector has a natural supersymmetric fermionic completion.  It
just so happens that the extended domain wall reduction does not yield
this natural completion, and instead gives rise to the inconsistent set
of fermions given by (\ref{eq:d54icf}).  The fields that we have
identified, namely $\psi_\mu^{(-)}$, $\lambda^{(+)}$ and $B_\mu$, is
suggestive of a vector multiplet coupled to a spin-$\fft32$ multiplet
(which is indeed the case for the ${\cal N}=2$ theory in five dimensions).  So
from this point of view, there is in fact a missing second vector that
would complete this coupled set of multiplets.  Presumably this missing
vector may be identified as an appropriate linear combination of $\hat
g_{\mu y}$ and $\hat A_\mu$.  Denoting this vector field strength as
$K_{\mu\nu}$, we speculate that it would correct the transformations,
(\ref{eq:d54icf}), so that $G_{\mu\nu}$ would be replaced by, for example,
$G_{\mu\nu}+K_{\mu\nu}$ and $G_{\mu\nu}-K_{\mu\nu}$ in the spin-$\fft32$
and spin-$\fft12$ equations, respectively.  If this were the case, the
additional matter would separate cleanly into independent
supermultiplets.

\section{Consistent Reduction of Scalar Potential and the Radion}

    So far we have considered brane-world reductions involving
$p$-forms, and a dilaton $\hat\phi$ with a running potential.  The
consistency of these reductions requires turning on the radion, and
equating the dilatonic and radionic degrees of freedom.  In this
section, we show that we can also obtain consistent brane-world
reductions for systems comprising a set of additional scalars as well
as the dilaton.  Specifically, our starting point is the
$(D+1)$-dimensional theory described by the Lagrangian
\be
\hat {\cal L} = \hat R\, {\hat *\oneone} - \ft12{\hat * d\hat\phi}\wedge
d\hat \phi -\ft12 {\hat *d\hat\Phi_i}\wedge d\hat\Phi_i + 
g^2\, e^{a\,\hat \phi}\, V(\hat\Phi_i)\,,\label{extrascal}
\ee
where $\Phi_i$ denotes an additional set of scalars, with a potential
$V(\Phi_i)$ that has a stationary point $V_0$.  Clearly, the solution
admits a domain wall solution (\ref{domainwall}) with $m^2=-\ft12
\Delta\, g^2\, V_0$.  We find that the scalars $\hat\Phi_i$ can be
consistently reduced into this domain wall world-volume spacetime,
provided that $a=\sqrt{2/(D-1)}$, corresponding again to $\Delta=-2$.  The
reduction ansatz is given by
\bea
d\hat s^2 &=& W^{-\ft{2}{D-1}}\, e^{2\alpha\, \varphi}\, ds^2 +
W^{-\ft{2D}{D-1}}\, e^{-2(D-2)\,\alpha\,\varphi}\, dy^2\,,\nn\\
e^{a\,\hat \phi} &=& W^{-\ft{2a^2}{\Delta}}\, e^{2(D-2)\,\alpha\,
\varphi}\,,\qquad \hat\Phi_i=\Phi_i\,.
\eea
We find that the resulting Lagrangian of the lower-dimensional theory
is given by
\be
{\cal L} = R\, {*\oneone} - \ft12{*d\varphi}\wedge
d\varphi -\ft12 {*d\Phi_i}\wedge d\Phi_i + 
g^2\, e^{2(D-1)\,\alpha\,\varphi}\, (V(\Phi_i)-V_0)\,,
\ee
where $\alpha^2=1/(2(D-2)(D-1)^2)$.  Thus we see that the coupling of
the dilaton $\phi$ in the lower dimensional scalar potential is of the form
$\sqrt{2/(D-2)}$, again corresponding to $\Delta=-2$.

  Scalar potentials of the type appearing in (\ref{extrascal}), with
a dilaton coupling $\Delta=-2$ in the scalar potential,
typically arise in gauged supergravities, such as those that come from
$S^3$ reductions of ungauged supergravities, where the
higher-dimensional 3-form field strength is taken to be proportional
to the volume of the $S^3$.

   General $S^3$ reductions of this kind were derived in
\cite{cvlupos3}.  Starting in $D$ dimensions from the Lagrangian
\be
\hat {\cal L}_D = \hat R\, {\hat *\oneone} -\ft12 {\hat *d\hat \phi}
\wedge d\hat \phi -\ft12 
e^{-b\, \hat \phi} \, {\hat *\hat F_\3}\wedge \hat F_\3\,,
\label{ungaugedlag}
\ee
where $b^2 = 8/(D-2)$, it is shown in \cite{cvlupos3} that the following
Kaluza-Klein ansatz gives a consistent $S^3$ reduction:
\bea
&&d\hat s_D^2 = Y^{\fft1{D-2}}\, \Big( \Omega^{\fft2{D-2}}\, ds_{D-3}^2
+ g^{-2}\, \Omega^{-\fft{D-4}{D-2}}\,  T^{-1}_{ij}\,
\cD\mu^i\, \cD\mu^j\Big) \,,\label{metans}\nn\\
&&e^{\sqrt{(D-2)/2}\, \hat\phi} = \Omega^{-1}\, Y^{(D-4)/4}\,,
\label{phians}\nn\\
&&\hat F_\3 = F_\3 + \ft16\, \ep_{i_1 i_2 i_3 i_4}\,
\Big( g^{-2}\, U\, \Omega^{-2}\,
  \cD\mu^{i_1}\wedge \cD\mu^{i_2} \wedge \cD\mu^{i_3}\,
\mu^{i_4} \label{fans}\\
&&\quad- 3g^{-2}\, \Omega^{-2} \,
D\mu^{i_1} \wedge \cD\mu^{i_2}\wedge \cD T_{i_3 j}\,
T_{i_4 k}\, \mu^j\, \mu^k  - 3g^{-1}\, \Omega^{-1}\, F_\2^{i_1 i_2} \wedge
\cD\mu^{i_3}\, T_{i_4 j}\, \mu^j \Big)\,,\nn
\eea
where  
\bea
&&\mu^i \, \mu^i = 1\,,\qquad \Omega = T_{ij}\, \mu^i\, \mu^j\,,\qquad
U = 2 T_{ik}\, T_{jk}\, \mu^i\, \mu^j - \Omega \, T_{ii}\,,\nn\\
&& Y = \det(T_{ij})\,,\label{somedefs}
\eea
and the indices $i,j,\ldots$ range of 4 values.  Here, 
a summation over repeated $SO(n+1)$ indices is understood.
The gauge-covariant exterior derivative $D$ is defined so that
\be
\cD\mu^i = d\mu^i + g\, A_\1^{ij}\, \mu^j\,,\qquad
\cD T_{ij} = dT_{ij} + g\, A_\1^{ik}\, T_{kj} + g\, A_\1^{jk}\, T_{ik}\,,
\label{gaugecov}
\ee
where $A_\1^{ij}$ denotes the $SO(4)$ gauge potentials coming from the
isometry group of the 3-sphere, and
\be
F_\2^{ij} = dA_\1^{ij} + g\, A_\1^{ik}\wedge A_\1^{kj}\,.
\label{fieldstrength}
\ee
Thus the lower-dimensional fields appearing in the Kaluza-Klein Ansatz
comprise the metric $ds_{D-3}^2$, the six gauge potentials $A_\1^{ij}$
of $SO(4)$, the ten scalar fields described by the symmetric tensor
$T_{ij}$, and the 2-form potential $A_\2$, whose (Chern-Simons
modified) field strength is $F_\3$.  The resulting $(D-3)$-dimensional
equations of motion can be derived from the Lagrangian \cite{cvlupos3}
\bea
{\cal L}_{D-3} &=& R\, {*\oneone} - \ft12
{*d\phi}\wedge d\phi - \ft14 \wtd T^{-1}_{ij}\, {*\cD \wtd T_{jk}}\wedge
\wtd T^{-1}_{k\ell}\, \cD\wtd T_{\ell i} \nn\\
&& - \ft12 Y^{-1}\, {*F_\3}\wedge
F_\3 -\ft1{4}\, Y^{-\fft12}\,  \wtd T^{-1}_{ik}\,
\wtd T^{-1}_{j\ell}\, {* F_\2^{ij}}\wedge F_\2^{k\ell}
-V\, {*\oneone}\,,\label{dm3lag}
\eea
where $Y\equiv \exp(-\sqrt{8/(D-5)}\, \phi)$, $\wtd T_{ij} \equiv 
Y^{-1/4}\, T_{ij}$ (implying that $\det (\wtd T_{ij})=1$), and 
the potential $V$ is given by
\be
V = \ft12 g^2\, Y^{\fft12}\, \Big(2 \wtd T_{ij}\, \wtd T_{ij}
  - (\wtd T_{ii})^2 \Big)\,.\label{s3pot}
\ee
The 3-form field strength $F_3$ is given by
\be
F_\3 = dA_\2 +\ft18 \ep_{ijk\ell}\, (F_\2^{ij}\wedge A_\1^{k\ell}
-\ft13 g\, A_\1^{ij}\wedge A_\1^{km}\wedge A_\1^{m\ell})\,.
\ee
The scalars $\wtd T_{ij}$ parameterise the coset $SL(4,\R)/SO(4)$.

   Focusing first on the scalar sector of this theory, we see that
there is a dilaton $\phi$ with exponential coupling in the scalar
potential that is exactly of the strength $\Delta=-2$ that we required
for our consistent brane-world reduction in this section.  The nine
scalars described by the unimodular symmetric tensor $\wtd T_{ij}$
correspond to the scalars $\Phi_i$ in our earlier general discussion.
In fact we can see that the strengths of the dilaton coupling to $F_3$
and $F_\2^i$ in (\ref{dm3lag}) are also exactly what we found to be
necessary in section 2 in order to obtain consistent brane-world
reductions of these fields too.  Thus we can start from the theory
(\ref{ungaugedlag}) in $D$ dimensions, perform a consistent $S^3$
reduction to obtain (\ref{dm3lag}) in $(D-3)$ dimensions, and then
perform a further brane-world consistent reduction, ending up with a
theory in $(D-4)$ dimensions that comprises the metric, a radion
$\varphi$, nine scalars $\wtd T_{ij}$ parameterising $SL(4,\R)/SO(4)$,
a 3-form $F_\3$, and the six $SO(4)$ Yang-Mills fields $F_\2^i$.  The
lower dimensional Lagrangian is given by
\bea
{\cal L}_{D-4} &=& R\, {*\oneone} - \ft12
{*d\varphi}\wedge d\varphi - \ft14 \wtd T^{-1}_{ij}\,
{*\cD \wtd T_{jk}}\wedge
\wtd T^{-1}_{k\ell}\, \cD\wtd T_{\ell i} \nn\\
&& - \ft12 e^{-\sqrt{\ft{8}{D-6}}\,\varphi}\, {*F_\3}\wedge
F_\3 -\ft1{4}\, e^{-\sqrt{\ft{2}{D-6}}\,\varphi}\,  \wtd T^{-1}_{ik}\,
\wtd T^{-1}_{j\ell}\, {* F_\2^{ij}}\wedge F_\2^{k\ell}
-V\, {*\oneone}\,,\label{dm4lag}
\eea
where the scalar potential $V$ is given by
\be
V =  \ft12 g^2\, e^{\sqrt{\ft{2}{D-6}}\, \varphi}\, 
\Big(2 \wtd T_{ij}\, \wtd T_{ij}   - (\wtd T_{ii})^2 + 8\Big)\,.
\ee

    It is interesting to note that we have started from the theory in
$D$ dimensions described by the Lagrangian (\ref{ungaugedlag}), where
there is no gauging and no Yang-Mills fields, and we have ended up in
$(D-4)$ dimensions with the theory described by (\ref{dm4lag}), where
we have $SO(4)$ Yang-Mills fields, and a set of scalars with a
non-trivial scalar potential.  However, remarkably, the scalar
potential in the $(D-4)$-dimensional gauged theory admits Minkowski
spacetime as a solution.  It is quite unusual in Kaluza-Klein
reductions that one can end up with non-abelian gauge theories coming
from a reduction on a compact space with isometries, and yet still
have a Minkowski vacuum solution.

     It is of interest therefore to examine the geometry of the
4-dimensional internal spaces that combine the $S^3$ reduction and the
brane-world reduction.  This can be seen most clearly by just looking
at the ``vacuum'' solution, where the $S^3$ is undistorted and the
solution in $(D-3)$-dimensional gauged theory is taken to be the
domain wall.  Retracing the steps described above, we see therefore
that this domain-wall solution lifts back to give
\be
d\hat s_D^2 = W^{-\ft{2}{D-2}}\, dx^\mu\, dx_\mu + 
W^{-\ft{2(D-1)}{D-2}}\, dy^2 + g^{-2}\, W^{-\ft{2}{D-2}}\, d\Omega_3^2
\,.
\ee
If we define a new coordinate $r$ in place of $y$, by setting $W=g^2\, r^2$,
the $D$-dimensional metric becomes
\be
d\hat s_D^2 = (g\, r)^{\ft{4}{D-2}}\, dx^\mu\, dx_\mu  
+ (g\, r)^{-\ft{2(D-4)}{D-2}}\, (dr^2 + r^2\, d\Omega_3^2)\,.
\label{d5brane}
\ee
(Here we have used the relation $m^2=4g^2$, which follows from the relation
given below (\ref{extrascal}).)  The metric (\ref{d5brane}) can be recognised
as the near-horizon limit of the $(D-5)$-brane in $D$ dimensions:
\be
d\hat s_D^2 = \Big(1 + \fft{Q}{r^2}\Big)^{-\ft{2}{D-2}}\, dx^\mu\, dx_\mu + 
+
\Big(1 + \fft{Q}{r^2}\Big)^{\ft{D-4}{D-2}}\, (dr^2 + r^2\, d\Omega_3^2)\,,
\label{d5brane2}
\ee
where $Q=g^{-2}$. 

   The ``internal'' four-dimensional metric in the vacuum solution
(\ref{d5brane}) can be seen to be singular at $r=0$, on account of the
conformal factor $(g\, r)^{-2(D-4)/(D-2)}$ that multiplies the 
flat 4-metric $dr^2 + r^2\, d\Omega_3^2$.  (An exception, of course, arises
if $D=4$, since then the entire metric is ``internal,'' and is merely the
Euclidean metric itself.)  Another way to view the internal geometry is
by introducing a new radial coordinate $\rho$, related to $r$ by
$g\, \rho = (g\, r)^{2/(D-2)}$.  The metric (\ref{d5brane}) then becomes
\be
d\hat s_D^2 = (g\, \rho)^2\, dx^\mu\, dx_\mu + \ft14 (D-2)^2\, d\rho^2 
+ \rho^2\, d\Omega_3^2\,.
\ee
Again, we see that the internal 4-metric is in general singular, 
describing a cone over $S^3$, with $D=4$ being the exceptional case
where the metric becomes non-singular.

    Finally, we remark that the ``anomaly term'' term ${\cal L}_{\rm anom} 
= -(D-26)\, m^2\, e^{\sqrt{2/(D-2)}\, \phi}$ in the $D$-dimensional
non-critical string effective action also has a $\Delta=-2$ coupling
strength.  It is easy to verify that one can perform a consistent
brane-world reduction on the theory described by (\ref{ungaugedlag}) 
together with the additional contribution ${\cal L}_{\rm anom}$.  A
particularly interesting case is the 27-dimensional non-critical bosonic
string, which can then be reduced to give a critical string in 
26 dimensions by means of this brane-world reduction. 

\section*{Acknowledgements}

    We are grateful to Ergin Sezgin for useful discussions.

\bigskip\bigskip

\centerline{\Large\bf APPENDICES}
\appendix

\section{Conventions for differential forms}

   Our conventions for differential forms are as follows.  A $p$-form
$\omega$ has components $\omega_{\mu_1\cdots \mu_p}$ such that
\be
\omega =\fft1{p!}\, \omega_{\mu_1\cdots \mu_p}\, dx^{\mu_1}\wedge 
\cdots \wedge dx^{\mu_p}\,.
\ee
The Hodge dual in $n$ dimensions is defined by
\be
{*(dx^{\mu_1}\wedge \cdots \wedge dx^{\mu_p})} = 
\fft{1}{(n-p)!}\, \ep_{\nu_1\cdots \nu_{n-p}}{}^{\mu_1\cdots \mu_p}\, 
dx^{\nu_1}\wedge \cdots \wedge dx^{\nu_{n-p}}\,,
\ee
which implies that the components of the dual ${*\omega}$, defined by
\be
{*\omega}= \fft1{(n-p)!}\, ({*\omega})_{\nu_1\cdots \nu_{n-p}}\, dx^{\nu_1} 
\wedge \cdots \wedge dx^{\nu_{n-p}}\,,
\ee
 are given by
\be
({*\omega})_{\nu_1\cdots \nu_{n-p}} = \fft1{p!}\, \ep_{\nu_1\cdots 
\nu_{n-p}}{}^{\mu_1\cdots \mu_{p}}\, \omega_{\mu_1\cdots \mu_p}\,.
\ee
We therefore have that
\be
{*\omega}\wedge \omega = \fft1{p!}\, \omega_{\mu_1\cdots \mu_p}\, 
\omega^{\mu_1\cdots \mu_p}\, {*\oneone}\,,
\ee
where 
\be
{*\oneone} = \sqrt{-g} \, dx^0\wedge dx^1 \wedge \cdots dx^{n-1}
\ee
is the volume form, and we are taking $\ep_{012\cdots} =+\sqrt{-g}$.  
Thus a Lagrangian with fields normalised so that
\be
e^{-1}\, L = R - \ft12 (\del\phi)^2 - \fft1{2 p!}\, 
F_{\mu_1\cdots \mu_p}\, F^{\mu_1\cdots \mu_p}\,,
\ee
can be written instead in terms of the $n$-form
\be
{\cal L} = R\, {*\oneone} -\ft12 {*d\phi}\wedge d\phi - \ft12
{*F}\wedge F\,.
\ee

\section{$S^2$ reduction of gauged $D=7$ supergravity}
\label{s2red}

      In this appendix, we address the question of whether the
$D=5$, ${\cal N}=2$ gauged
supergravity coupled to a vector multiplet discussed in section
\ref{5to4susy} can also be obtained from an $S^2$ reduction of the $D=7$,
${\cal N}=2$ gauged pure supergravity discussed in \ref{7to6red1}.

    At the level of the bosonic sector, the following
ansatz, which was considered in general in \cite{instanton}, 
gives a consistent reduction of the seven-dimensional
gauged supergravity (\ref{d7lag-2}) to $D=5$:
\bea
ds_7^2&=& e^{-\ft{2}{5}\sqrt{\ft23}\,\phi}\, ds_5^2 +
          \fft{2}{\lambda^2}\,e^{\ft35\sqrt{\ft23}\,\phi}\, d\Omega_2^2
\,,\qquad \hat \phi = \sqrt{\fft35}\,\phi\,\nn\\
\hat F_\4&=&\fft2{\lambda^2}F_\2\wedge\Omega_\2\,,\qquad
\hat F^3_\2=\fft2\lambda\,\Omega_\2 + G_\2\,,\qquad
\hat F^1_\2=0=\hat F^2_\2\,.\label{7to5ans}
\eea
The resulting five-dimensional equations of motion are those of the bosonic
sector of five-dimensional
gauged supergravity coupled to a vector multiplet, described by
(\ref{d5lag}) with $g_2=0$ and 
\be
g_1^2=2g^2 + \ft12\lambda^2\,.\label{squared}
\ee
Note that the lower-dimensional ``cosmological constant'' $g_1^2$ is
the sum of contributions from both $\lambda$ and $g$.  This implies
that there is a 1-parameter family of ways of obtaining the same
five-dimensional bosonic theory, with different proportions of the
contributions to the lower-dimensional cosmological term coming from
the curvature of the reduction 2-sphere versus the already-present
cosmological term in the seven-dimensional gauged theory.  In fact in
one extreme, we can take $\lambda=0$, which amounts to making a $T^2$
reduction of the $D=7$ gauged theory.  At the other extreme, we can
take $g=0$, which amounts to an $S^2$ reduction of the {\it ungauged}
$D=7$ theory.

    The situation changes somewhat when we include the fermionic sector
in the reduction.   We find that for the general case with $g$ and $\lambda$ 
both non-zero, we cannot obtain a consistent five-dimensional supersymmetric
theory.  In fact this can be illustrated by considering a simple known
supersymmetric solution of the five-dimensional gauged theory, namely
the domain wall that preserves half of the $D=5$ supersymmetry.  In
$D=5$ this is given by
\bea
ds_5^2 &=& W^{-\ft23}\, dx^\mu\, dx_\mu + W^{-\ft83}\, dy^2\,,\nn\\
e^{\phi} &=& W^{-\sqrt{\ft23}}\,,
\eea
where $W=1+ m\, y$ and $m=g_1$.
Lifting this to $D=7$ using (\ref{7to5ans}), we find
\bea
d\hat s_7^2 &=& W^{-\ft25}\, (dx^\mu\, dx_\mu + \fft{2}{\lambda^2}\, 
d\Omega_2^2) + W^{-\ft{12}5}\, dy^2\,,\nn\\
e^{-\ft1{\sqrt{10}}\hat \phi} &=& W^{\ft15}\,,\qquad \hat F_\2^3
= \fft2{\lambda}\, \Omega_\2\,.\label{d7soln}
\eea

   To illustrate the issue of supersymmetry, it suffices to consider
just the dilatino transformation rule given in (\ref{eq:7fsusy}).  
In the background (\ref{d7soln}), the dilatino variation
gives
\be
\delta\hat \lambda_i = \ft{1}{\sqrt{10}}\, W^{\ft15}\, \Big[
\ft1{\sqrt{2}}\, m\, \hat \gamma^5\, \epsilon_i + g\, \epsilon_i +
\ft{i}{2}\, \lambda\, (\sigma_3)_i{}^j\, \hat \gamma^{67}\, \epsilon_j
\Big]\,,
\ee
where the $y$ direction is denoted by ``5'' and the two directions 
on $S^2$ are denoted by ``6'' and ``7.''  We see that to get preserved 
supersymmetry, we must have
\be
m = \sqrt2\, (\pm g \pm \ft12 \lambda)\,.
\label{eq:dwrval}
\ee
This is compatible with (\ref{squared}) only if $g=0$ or $\lambda=0$.

In general, one may search for a fermion reduction by substituting the
bosonic ansatz, (\ref{7to5ans}), into the supersymmetry transformations
(\ref{eq:7fsusy}).  For the dilatino, we find
\begin{eqnarray}
\delta\hat\lambda_i&=&-i\sqrt{\ft65}e^{\fft15\sqrt{\fft23}\phi}
\Bigl[-\ft{i}4\gamma^\mu\partial_\mu\phi\delta_i{}^j-\ft1{8\sqrt{3}}
\gamma^{\mu\nu} \Bigl(e^{\fft12\sqrt{\fft23} \phi}G_{\mu\nu}(\sigma^3)_i{}^j
+\sqrt{2}e^{-\sqrt{\fft23}\phi}F_{\mu\nu}(i\gamma^{67})\delta_i{}^j
\Bigr)\nn\\
&&\qquad+\ft{i}{2\sqrt{6}}\Bigl(\sqrt{2}g+\ft1{\sqrt{2}}\lambda
(i\gamma^6\gamma^7)(\sigma^3)_i{}^j\Bigr)e^{-\fft12\sqrt{\fft23}\phi}
\Bigr]\hat\epsilon_j\,,
\end{eqnarray}
which may be compared with the corresponding $D=5$ dilatino transformation
of (\ref{eq:5susy}).  A more direct comparison may be obtained by
converting the spinors into their natural five-dimensional counterparts,
following (\ref{eq:d75gam}).  However, even at present, it is clear that
for the reduced dilatino to transform properly, one requires (up to an
overall sign)
\begin{equation}
m=\sqrt{2}[g+\ft12\lambda(i\gamma^6\gamma^7)(\sigma^3)]
\label{eq:linspot}
\end{equation}
(where this notation refers to the eigenvalues of the Dirac matrices on
the reduced spinor parameter $\epsilon$).  This confirms
(\ref{eq:dwrval}), and furthermore indicates that the sign is dependent
on the orientation of the $S^2$.

The supersymmetry of several breathing
mode sphere reductions was considered in Ref.~\cite{Liu:2000gk}.  There
it was conjectured that breathing mode reductions could be consistent,
at least in the case of reduced supersymmetries.  In the examples of
Ref.~\cite{Liu:2000gk}, contributions to the lower-dimensional potential
were similarly quadratic, as in (\ref{squared}), while contributions to
the ``superpotential'' were linear, as in (\ref{eq:linspot}).  However,
unlike the present case, in those examples, the potential and
superpotential comprised more than a single exponential, thus allowing
the functional relationship between potential and superpotential to be
satisfied for arbitrary values of the parameters.  Furthermore, the
earlier examples were all for sphere reduction of ungauged
supergravities.  From this point of view, the obstruction to having a
consistent reduction on the fermions is similar to that discussed for
the bosonic examples of section~\ref{sec:ebe}.  Namely, turning on the
graviphoton $\hat F_\2^3$ (even when restricted to the sphere direction)
gives rise to inconsistencies in the fermion sector.


\end{document}